\documentclass{agujournal2019}
\usepackage{url} 
\usepackage{lineno}

\usepackage{xcolor}
\usepackage{float}
\usepackage[final]{changes} 
\setaddedmarkup{\textcolor{blue}{\textbf{#1}}}
\setdeletedmarkup{\textcolor{red}{\sout{#1}}}

\usepackage{soul}
\usepackage{tabularx}
\usepackage{threeparttable}

\draftfalse

\journalname{Space Weather}

\begin{document}

%
%


\title{The CHASM-SWPC Dataset for Coronal Hole Detection \& Analysis}

%
%




\authors{
    C.~Beck\affil{1}\thanks{These authors contributed equally to this work.},
    E.~Smith\affil{1}$^\ast$,
    K.~Katuwal\affil{2},
    R.~Kafle\affil{1},
    J.~Whitehill\affil{1}
}


\affiliation{1}{Worcester Polytechnic Institute, 01609, Worcester, United States}
\affiliation{2}{New Mexico State University, 88003, Las Cruces, United States}




\correspondingauthor{Cutter Beck}{cjbeck@alum.wpi.edu}



\begin{keypoints}
\item CHASM enables the first digitization of NOAA SWPC synoptic coronal hole maps for ML applications
\item We release CHASM-SWPC datasets aligned with multi-channel SDO imagery and a trained coronal hole detection model
\item Training CHRONNOS on CHASM-SWPC improves accuracy, TSS, and IoU over training on SPoCA-CH labels
\end{keypoints}

%
%

%
%


\begin{abstract}

Coronal holes (CHs) are low-density solar coronal regions with open magnetic field lines \cite{cranmer_coronal_2009}. Since 1972, the National Oceanic and Atmospheric Administration (NOAA) Space Weather Prediction Center (SWPC) has produced daily drawings identifying CHs \cite{solar_synoptic_map}. Despite containing decades of expert annotations, these maps are not available in a format suitable for machine learning, and thus have not been used to train CH detection algorithms. We developed CHASM (Coronal Hole Annotation using Semi-automatic Methods), the first method to digitize the SWPC drawings into binary segmentation masks. The resulting masks constitute the CHASM-SWPC dataset, a high-quality dataset to train and test automated CH detection models, which is released with this paper. Based on state-of-the-art dense semantic segmentation algorithms \cite{kirillov_segment_2023}, the CHASM software tool enables users to rapidly and accurately annotate SWPC drawings. Post-processing of CHASM annotations aligns the binary segmentations with solar disk imagery provided by the Solar Dynamics Observatory (SDO). We then trained multiple CHRONNOS architecture \cite{jarolim_multi-channel_2021} neural networks using the CHASM-SWPC dataset and compared their performance. The CHASM tool enabled us to annotate 1446 SWPC drawings, filtered down to 1,112 drawings, making up our CHASM-SWPC-1112 dataset. Training CHRONNOS on the CHASM-SWPC-1112 dataset achieved an accuracy of 0.9802, a True Skill Statistic (TSS) of 0.6807, and an intersection-over-union (IoU) of 0.5668; this is higher than the original CHRONNOS model trained on SPoCA-CH, which achieved an accuracy of 0.9708, a TSS of 0.6749, and an IoU of 0.4805, when evaluated on the CHASM-SWPC-1112 test set at an optimized binary threshold for both.
\end{abstract}

\section*{Plain Language Summary}
Coronal holes are regions on the Sun that are known to stream charged particles into space, sometimes affecting the Earth by disrupting power grids, satellites, and degrading GPS. Accurately identifying coronal holes leads to better space weather predictions and mitigation of these effects. However, there has been a lack of high-quality data to train automated coronal hole detection models. Since 1972, experts at the National Oceanic and Atmospheric Administration (NOAA) Space Weather Prediction Center (SWPC) have produced daily drawings of coronal holes, but these drawings are not usable for machine learning in their current format. In this work, we developed the Coronal Hole Annotation using Semi-automatic Methods (CHASM) tool to convert these SWPC drawings into a machine learning-ready format for the first time. Using CHASM, we converted \replaced{1446}{1,111} of the drawings into the new format and aligned \replaced{1407 of those}{the results} with solar satellite imagery. With this new source of data, we trained a computer model to detect coronal holes in unseen images of the sun with higher accuracy than existing methods. The CHASM-SWPC dataset is based on expert knowledge and is publicly available, therefore it enables the creation of new detection models and supports future coronal hole research.

%
%

%


%
%
%
%

\section{Introduction}
\label{1_introduction}

\par Coronal holes are \replaced{lesser}{low}-activity regions characterized by their low density and open, unipolar magnetic fields \added{\protect\cite{1982SoPh...79..149H,2026ApJ...999...63K}} in the solar corona. Open magnetic field lines from coronal holes (CHs) are associated with the formation of fast solar wind that can carry charged particles toward the Earth, bombarding the magnetosphere and atmosphere \cite{cranmer_coronal_2009}. Without advance notice of when charged particles are expected to hit the Earth, satellite systems for communications and GPS can be degraded, astronauts may experience higher radiation, and power grids can be overwhelmed due to the inability to keep up with customer power demands \cite{schwenn_space_2006, noaa_scales}. By detecting and tracking CHs, scientists can provide better space weather predictions to these stakeholders.

\par
\begin{figure*}[t]
    \centering
    \includegraphics[width=1\linewidth]{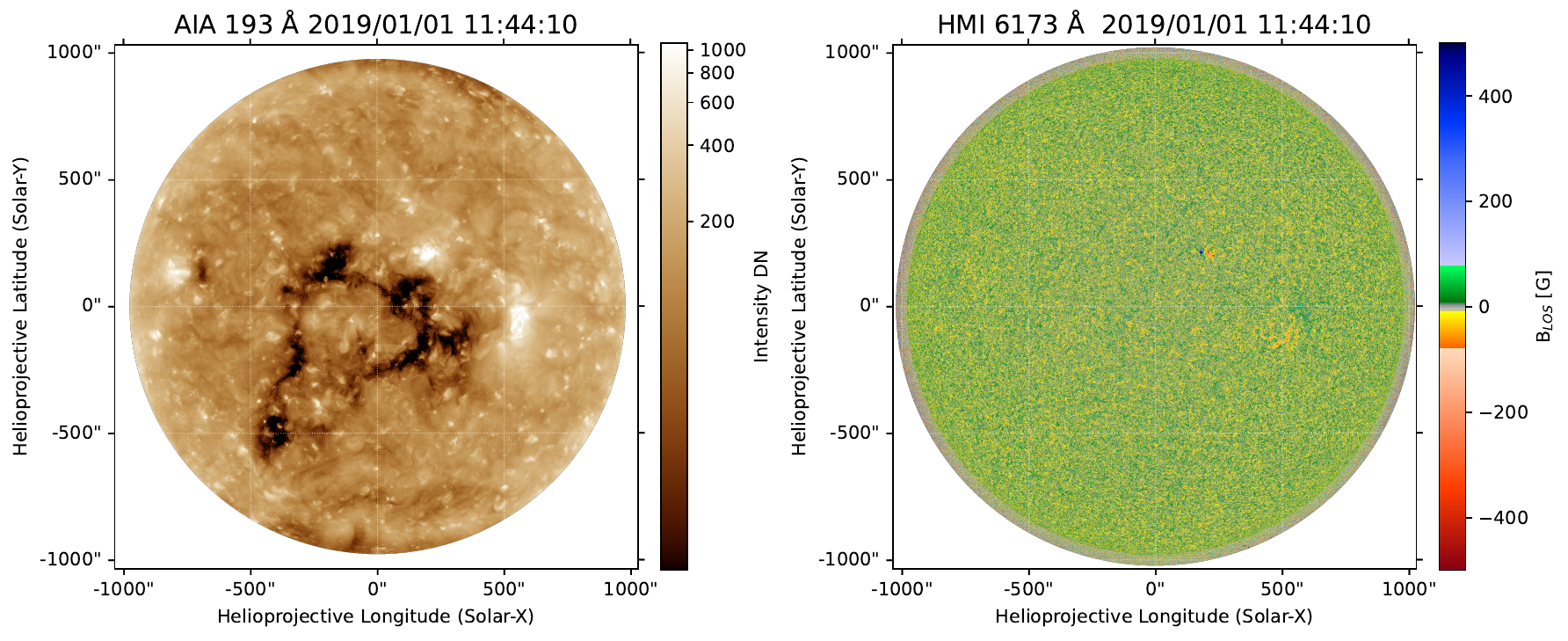}
    \caption{An EUV image of the Sun with a dark coronal hole in the center with intensity under 200 DN (digital number) (left) and the corresponding magnetogram (right) from 2019-01-01. Source: \cite{virtual_solar_observatory} with SunPy \cite{sunpy}.}
    \label{fig:intro_EUV_Mag_pair}
\end{figure*}

 In the extreme ultraviolet (EUV) spectrum, CHs appear as dark patches. In Figure \ref{fig:intro_EUV_Mag_pair},  a CH can be seen in 193\AA\space with its corresponding magnetogram taken by the Solar Dynamics Observatory’s (SDO) Atmospheric Imaging Assembly (AIA) and Helioseismic and Magnetic Imager (HMI) instruments, respectively. Notice that it is difficult to visually detect the CH in the magnetogram, whereas it is starkly contrasted as a dark patch on the light solar disk in 193\AA. In other EUV imagery of the sun at different wavelengths, solar features like solar filaments also appear as dark patches. Using isolated imagery of the sun leads to difficulty in differentiating CHs from other solar features and therefore, a multi-wavelength approach to detection is beneficial for more accurate reporting.

\par Daily since June 2, 1972, expert space weather forecasters at the National Oceanic and Atmospheric Administration (NOAA) Space Weather Prediction Center (SWPC) have \replaced{created hand drawings}{manually drawn synoptic maps} (see Figure \ref{fig:four-wide}, panel a) of the solar disk \cite{solar_synoptic_map} using imagery in the EUV and other spectra to identify solar features. This process is time-consuming, and the forecasters only create up to 2 drawings of the solar disk per day.  Automated detection methods that can create maps on demand are thus desirable. While early such methods relied on simple pixel-intensity thresholding of solar imagery, contemporary approaches are based on machine learning using  convolutional neural networks \cite{illarionov_segmentation_2018, jarolim_multi-channel_2021}. One bottleneck to further improving machine learning-based CH detectors is the lack of high-quality annotations, and hence pseudo-labels obtained from clustering algorithms such as SPoCA \cite{verbeeck_spoca-suite_2013} are sometimes used as a proxy. However, given that the NOAA SWPC has labeled CHs everyday for the past 50 years, an important opportunity is available to improve the training and testing datasets of automated CH detection methods.

\par In this paper, we pursue a semi-automated approach to efficiently convert SWPC \deleted{synoptic }drawings from their current hand-drawn format into a high-quality annotated dataset that can be readily used for training and testing machine learning models for CH detection. Our method harnesses state-of-the-art computer vision algorithms such as the Segment Anything Model (SAM) \cite{kirillov_segment_2023} for dense semantic segmentation, which assigns a class label to every pixel of the input image as seen in Figure \ref{fig:four-wide} (b). SAM can produce highly detailed object boundaries which are especially useful for irregularly shaped objects like CHs in the \replaced{SWPC drawings}{drawn synoptic maps}. Our tool also captures the SWPC\added{-labeled per} CH confidence value and assigned polarity from the \deleted{synoptic }drawing. Using CH \replaced{binary segmentation masks}{annotations} obtained from our semi-automated pipeline, we conduct machine learning experiments based on the CHRONNOS (Coronal Hole RecOgnition Neural Network Over multi-Spectral-data) architecture and show that training this network on our CHASM-SWPC data yields substantially different predictions compared to the same architecture trained on the standard SPoCA-CH labels used in \citeA{jarolim_multi-channel_2021}. In particular, \replaced{training on CHASM-SWPC produces predictions that are more consistent with SWPC expert annotations than training on SPoCA-CH pseudo-labels.}{training on CHASM-SWPC helps the network produce outputs that are more similar to the SWPC hand\replaced{ drawings}{-drawn synoptic maps}, compared to the identical neural network architecture trained on SPoCA-CH pseudo-labels}. We are releasing our CH annotation tool (CHASM: Coronal Hole Annotation using Semi-automatic Methods) as well as a dataset \added{(CHASM-SWPC)} of 1407 \replaced{CHASM annotations containing CH binary segmentation masks}{pre-segmented CH masks} with their \added{SWPC-}assigned \added{per CH} polarity\replaced{ and}{,} \deleted{SWPC }confidence value, \replaced{as well as}{and} corresponding EUV and magnetogram imagery \added{from the SDO}\deleted{(CHASM-SWPC)}.

\section{Related Work}
\label{2_related_work}

\subsection{Algorithmic Approaches to CH Detection}
\par Over the last 20 years, CH detection methods have combined and built on methods like intensity thresholding, pixel-wise clustering, and the modern use of neural networks. One classic approach, which is not based on machine learning, used intensity thresholding  \cite{henney_automated_2007,garton_automated_2017} by determining a specific value or values which denote the expected intensity of a CH in a given EUV image and filtering on that value to extract CHs. Another classic technique is pixel-wise clustering, which groups similar pixels based on a heuristic, such as in the K-Means algorithm \cite{k_means} or the possibilistic clustering algorithm outlined in \cite{verbeeck_spoca-suite_2013}. The SPoCA (Spatial Possibilistic Clustering Algorithm) uses fuzzy clustering to separately identify solar filaments and CHs.

Over the past 8 years, researchers have largely shifted to machine learning and  neural networks  for CH detection. In particular, \citeA{illarionov_segmentation_2018} uses the U-Net architecture \cite{ronneberger_u_net_2015} with just a 193\AA\space image as input. CHRONNOS \cite{jarolim_multi-channel_2021} uses the U-Net architecture and a progressive growing technique on 8 different input channels (7 EUV, 1 magnetogram) to achieve higher accuracy in locating CHs.

Most recently, \citeA{wang_solar_2025} present the Solar Activity AI Forecaster, which can detect both active regions and CHs. Their detector is based on the Vision Transformer neural network architecture \cite{dosovitskiy2020image} and is trained on \replaced{CH boundary labels their team created}{SWPC images that their team labeled for the CH boundaries} by manually referring to (rather than automatically processing) SWPC \replaced{drawings}{synoptic maps}. Hence, their labeling process likely generates less exact CH boundaries than our tool, which was based on a semantic segmentation system \cite{kirillov_segment_2023} that produces highly detailed boundaries. In addition, \citeA{wang_solar_2025} implemented an iterative semi-automatic labeling procedure whereby an initial set of annotations was used to train the network. The network was then used to predict solar features on a larger set of images, which human annotators verified. Finally, the verified predictions were added to the training pool, and the process was iterated. In contrast to our work, the dataset that they release contains only the \emph{predictions} of their trained neural network on the corresponding solar images, not the annotations of SWPC \replaced{drawings}{synoptic maps} themselves.

\subsection{Datasets}
\label{3_datasets}
While classic CH methods  based on thresholding or clustering do not require CH \deleted{boundary }labels for training, neural network-based methods require labeled datasets. Given the lack of high-quality CH annotations, machine learning models have  instead been trained on  pseudo-labels obtained from classic CH detection models such as SPoCA-CH \cite{DELOUILLE2018365},
which is an adapted version of the SPoCA \cite{verbeeck_spoca-suite_2013}  but also removes solar filaments. SPoCA-CH labels from \citeA{DELOUILLE2018365,jarolim_multi-channel_2021} are not provided as pixel-wise segmentations.

For testing, it has been common practice \cite{garton_automated_2017,illarionov_segmentation_2018} to visually compare the outputs of trained CH detectors to \replaced{hand drawings of the solar disk}{hand-drawn synoptic maps} \cite{solar_synoptic_map, kpvt_choles, moswoc}. Quantitative comparison has been hindered by the lack of digitized \replaced{solar disk drawings}{synoptic maps}.

\par The expert-annotated  SWPC \replaced{drawings}{synoptic maps} present an opportunity to improve the ground truth used by CH prediction models. These drawings are a wealth of expert information including CH boundaries, polarity, and a confidence value from the forecaster (1=uncertain to 4=good) \cite{solar_synoptic_map}. Digitizing these drawings for machine learning purposes is the focus of our work. 

\section{Annotation Tool}

\begin{figure*}[h!]
    \centering
    \includegraphics[width=1\linewidth]{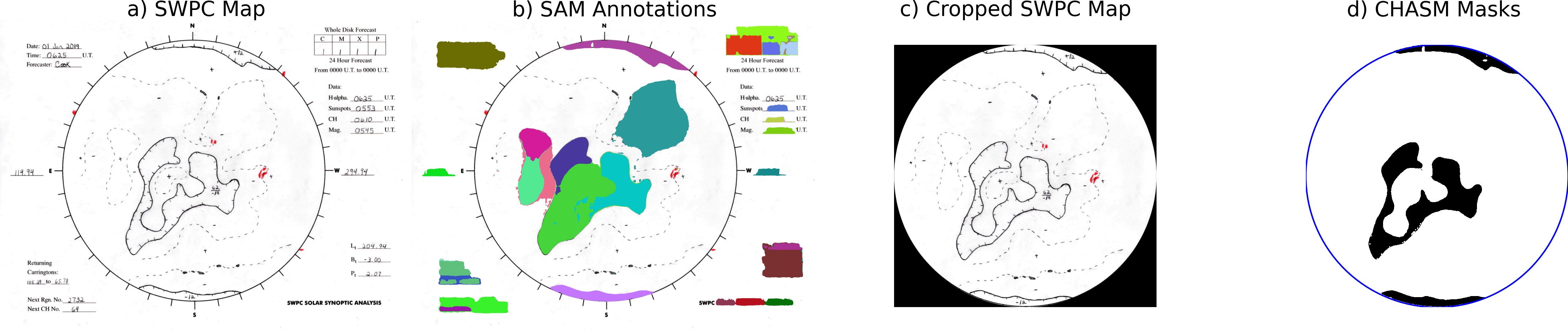}
    \caption{CH segmentation pipeline example for an observation made on 2019-01-01. From left to right: SWPC \replaced{drawing}{synoptic map} with \deleted{hand-drawn }CHs; SAM-generated segmentation masks overlaid on the SWPC \replaced{drawing}{map}; Cropped SWPC solar disk extracted using Hough circle detection; CHASM binary \replaced{segmentation masks}{mask} with CH pixels in black and the Hough circle in blue.}
    \label{fig:four-wide}
\end{figure*}

\par We developed CHASM, a Python-based software tool that allows users to quickly and accurately label CHs from SWPC \replaced{drawings}{synoptic maps} utilizing the Segment Anything Model (SAM) \cite{kirillov_segment_2023} segmentation masks. Once human labeling of the SAM masks is complete, an automatic post-processing step is employed to prepare the selections for automatic CH detection algorithms. The steps to turn the SWPC \replaced{drawings}{maps} into binary segmentation \replaced{masks}{maps} are as follows:

\begin{enumerate}
    \item Download \replaced{hand}{synoptic map} drawings from the SWPC \cite{solar_synoptic_map}.
    \item Automatically segment each drawing with SAM \cite{kirillov_segment_2023}.
    \item Manually select CHs using CHASM.
    \item Post-process selections to scale into appropriate masks.
\end{enumerate}

\subsection{Downloading SWPC Drawings}
SWPC drawings were downloaded automatically from the SWPC drawing repository \cite{solar_synoptic_map} using a Python script. Each drawing is then converted to JPEG format.

\subsection{Segmenting \replaced{SWPC}{Synoptic} Drawings with SAM}
From each SWPC \replaced{drawing}{image}, we extracted segmentation masks using Meta's SAM \cite{kirillov_segment_2023} in its ``everything'' segmentation mode. This generates numerous segmentation masks for all components of the \deleted{synoptic }drawing, including unwanted segmentations such as text and background, as seen in Figure \ref{fig:four-wide}(b). Based on pilot testing, we configured SAM's segmentation parameters to the following values: \path{pred_iou_thresh} = 0.80, \path{stability_score_thresh} = 0.80, \path{crop_n_layers} = 1, and \path{crop_n_points_downscale_factor} = 2. All other parameters were kept at their default values.

\subsection{Using the CHASM Tool}
\label{CHASMUse}
\par
\begin{figure*}[h!]
    \centering
    \includegraphics[width=\hsize]{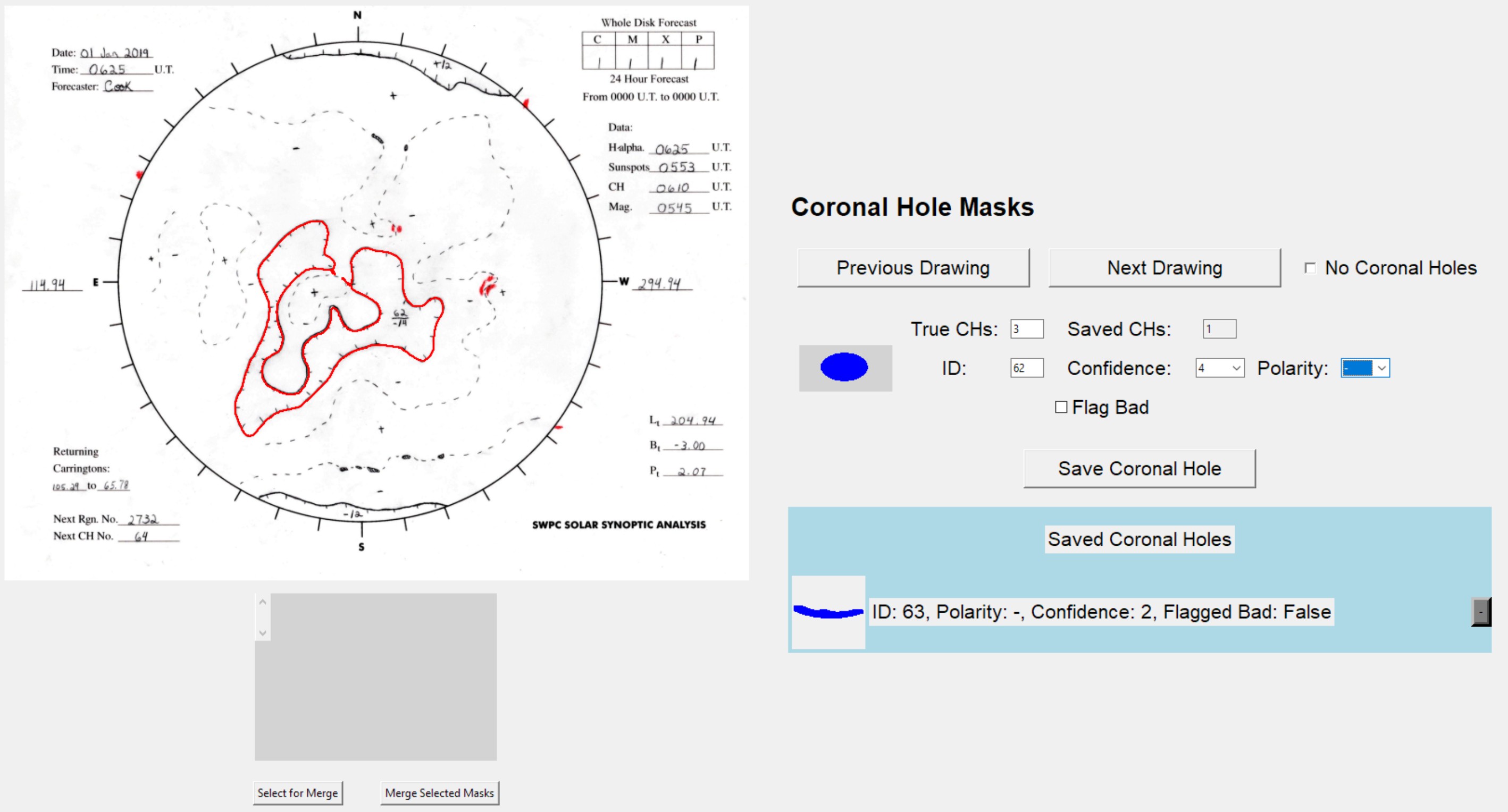}
    \caption{The CHASM tool to digitize SWPC \replaced{drawings}{maps}.}
    \label{fig:CHASM_Selection}
\end{figure*}

 After segmenting each SWPC \replaced{drawing}{synoptic map} with SAM, both the SWPC \replaced{drawing}{image} and the SAM segment\added{ation}s are loaded into CHASM. The user clicks the ``Next Drawing'' button seen in Figure \ref{fig:CHASM_Selection} to load the drawing image and segmentations; doing so displays all the available masks that can be clicked on for CH determination. The method of collecting a segmentation for a given day is as follows:

\begin{enumerate}
    \item Click on a CH to select the SAM segmentation mask.
    \item Modify the mask by merging with or subtracting from other masks to best fit the CH.
    \item Enter relevant information for the CH including \added{SWPC-assigned} confidence, ID, and if the mask is correct using the flag bad checkbox.
    \item Click ``Save CH''.
    \item Repeat steps 1-4 until all CHs have been cataloged for that day.
    \item Record the correct number of CHs in image.
    \item Click ``Next Drawing'' and repeat steps 1-5 until all days in batch have been collected.
\end{enumerate}

\par A user can click on any of the masks to see the underlying \replaced{SWPC drawing}{image} with the \replaced{CH}{mask} boundary visualized in red, as in Figure \ref{fig:CHASM_Selection}. If there appears to be multiple possible masks for one CH, the user can cycle through the masks in that area using the keyboard. For this particular example, the user also registered the SWPC CH ID as 62, the \added{SWPC-assigned} confidence as 4, and the polarity as negative (``-'') in the proper fields on the right. The user also noted under True CHs the presence of 3 CHs in the image; this information is later used to filter out days containing a CH that is missing a SAM mask.

\par To improve accuracy of the annotated CH boundaries, CHASM provides features to allow the user to compute the union of two SAM masks as well as the difference between two masks. In practice, we found this ``editing''  to be very important for harnessing as much data as possible. The end result was an annotation process with \replaced{highly consistent CH segmentation in most cases}{near-perfect CH segmentation}.

\subsection{\deleted{Mask }Post-Processing \added{CHASM Annotations}}

\par  CHASM outputs \replaced{annotations}{data} consisting of a list of individual CH binary segmentation masks, whose \replaced{dimensions}{sizes} equal those of the original SWPC \replaced{drawings}{synoptic maps}, along with associated metadata (polarity, confidence, and flag as ``bad''). Because the \replaced{drawings}{raw synoptic maps} are rectangular (they contain the solar disk as well as various text labels; see upper-left of Figure \ref{fig:CHASM_Selection}), the next step is to  crop and align these masks with the \replaced{square}{4096$\times$4096 pixel} solar disk images from the \replaced{SDO/AIA and SDO/HMI}{SDO AIA/HMI}
\added{
The following process is used to identify the solar disk in the SWPC drawings and crop the CH segmentation masks in the CHASM annotations to that disk:
}

\begin{enumerate}
    \item Use the Hough transform to perform circular edge detection on the solar disk in the SWPC drawing \cite{duda_use_1972}
    \item Use the center of the identified circle as the center point in each CH segmentation mask and crop the rectangular mask to a square with side length $2r_\mathrm{c}$, where $r_\mathrm{c}$ is the radius of the Hough circle in pixels
    \item Merge all cropped CH segmentations into one square binary segmentation mask containing all CHs in the SWPC drawing
\end{enumerate}

\added{
SWPC drawings list extra information outside of the solar disk, including the B-angle ($-7.23^\circ \le B_t \le 7.23^\circ$), the angle between the Earth's current position in its tilted orbit and the Sun's equator, and position angle ($-26.31^\circ \le P_t \le 26.31^\circ$), the angle between the Earth's geocentric north pole and the Sun's rotational north pole \protect\cite{solar_synoptic_map}. SWPC drawings are oriented with solar north at the top, therefore, $P_t$ is inherently corrected for by this convention. $B_t$ is not explicitly corrected for in the drawings. We do not correct for $B_t$ ourselves since the SWPC drawings and SDO images we pair to them are sufficiently close in time that they share approximately the same $B_t$. Due to our download tolerance (outlined in Section \protect\ref{SDOImagery}), the SDO imagery timestamp may precede the SWPC drawing timestamp by up to one hour.
}

\added{
The resulting square binary segmentation mask cannot be directly aligned with the SDO imagery as the imagery has not had a $P_t$ correction and $r_\mathrm{c} \neq r_\mathrm{o}$, where $r_\mathrm{o}$ is the observed radius of the solar disk in pixels in the SDO imagery. The observed radius of the solar disk changes as the Earth orbits the Sun from the SDO's perspective in geosynchronous orbit \protect\cite{sdo_pdf}, so image processing must be applied to normalize the images over time. SDO images are downloaded as 512$\times$512 pixel square images and have the same normalization process applied as outlined in \protect\cite{jarolim_multi-channel_2021}. Key steps of the normalization procedure include: center and rotate the solar disk with solar north at the top, scale the image to have a field of view of 1.1$r_\mathrm{o}$ (the solar disk with 0.1$r_\mathrm{o}$ padding around it), and for HMI images, all pixels off of the solar limb are set to zero. No other limb corrections are applied. To align the SWPC solar disk to the SDO solar disk, we must scale down the larger square binary segmentation mask so that the solar disks have the same radius and padding.
}

\added{
The observed radius of the solar disk in arcseconds, $r_\mathrm{oa}$, and the resolution in arcseconds per pixel, $\theta$, are provided as headers in the SDO image metadata. We can calculate $r_\mathrm{o} = r_\mathrm{oa} / \theta$. Given $r_\mathrm{o}$, $r_\mathrm{c}$ can be scaled down such that $r_\mathrm{c} = r_\mathrm{o}$. With the square binary segmentation mask scaled down to have side length now equal to $2r_\mathrm{o}$, the mask can be padded with zeros to increase the side length to $2.2r_\mathrm{o}$, leaving $0.1r_\mathrm{o}$ pixels around the solar disk not labeled as CHs.
}

\added{
When the SWPC creates the drawings, the products used to identify key features on the solar disk may differ slightly in their observation times. This means the observation used to identify CHs, for example, may differ by a few minutes from the observation used to identify active regions. This and our download tolerance do not have a large effect on matching SWPC drawings to the SDO imagery, based on visual registration qualtiy checks we performed throughout our experimentation.
}

\deleted{To detect the circular boundary of the solar disk -- whose position can vary slightly in different synoptic maps -- we use a Hough transform for edge detection \protect\cite{duda_use_1972}}

\deleted{SDO metadata provides the dimension of the solar radius, Ro, in a given image in pixels. The apparent size of Ro changes as the Earth orbits the Sun from the SDO's perspective in geosynchronous orbit \protect\cite{sdo_pdf}. It is important to use Ro from each image to reduce sizing errors when aligning them with CHASM segmentation masks.}

\section{CHASM-SWPC Dataset}

\par Using the CHASM tool, our labeling team (the first two authors of this manuscript) annotated a total of 1446 SWPC \replaced{drawings}{synoptic maps}: 359 from 2017, 365 from 2019, 359 from 2022, and 363 from 2024. These particular years were chosen to represent different portions of the solar cycle and to obtain a comprehensive and temporally consistent data set including solar activity variations. Collecting data from 2017 and 2019 ensures we capture declining and minimum solar activity, and the data from 2022 and 2024 captures rising and maximum solar activity.  The total annotation time, summed over the team, was about 14 hours.

\subsection{AIA and HMI Images}
\label{SDOImagery}

\par The CHRONNOS architecture requires seven EUV wavelength images from SDO/AIA, one line-of-sight (LoS) magnetogram from SDO/HMI, and one ground truth CH binary segmentation mask as inputs for training. 
The filename of each SWPC \deleted{synoptic map }drawing provides a date and timestamp referring to the time the observation was made, and we used these \added{dates} to find the AIA and HMI images that matched the \replaced{drawing}{synoptic map} as closely as possible in time. In general, AIA images are taken every 12 seconds. However, when we downloaded images from the \citeA{jsoc}, we discovered that some images were not available for our specified timestamps. To maximize the number of images that we could include in the dataset, we \deleted{thus }allowed a tolerance between the time \replaced{listed on the drawing}{of the synoptic map} and the corresponding satellite image. We chose a matching tolerance of either 1 hour or 1 minute \added{before the SWPC drawing for the SDO imagery}, resulting in different datasets, as described below. Note that these are both more stringent than the 6 hour tolerance implemented by CHRONNOS \cite{jarolim_multi-channel_2021}. After downloading each image, we applied the SDO image pre-processing methods outlined in \citeA{jarolim_multi-channel_2021}. 

\begin{table}[h!]
\caption{Dataset Counts}
\label{tab:DatasetCounts}
\centering
\small
\begin{tabular*}{\linewidth}{@{\extracolsep{\fill}} l r r r r r}
    \hline\hline
    \textbf{Description} & \textbf{2017} & \textbf{2019} & \textbf{2022} & \textbf{2024} & \textbf{Total} \\
    \hline
    SWPC Drawings & 359 & 365 & 361 & 363 & 1448 \\
    SAM Masks from SWPC Drawings & 359 & 365 & 361 & 363 & 1448 \\
    CHASM Annotations from SAM Masks, $S_0$ & 359 & 365 & 359 & 363 & 1446 \\
    \hline
    \quad SDO Image Sets missing at least 1 Wavelength, $M$ & 8 & 0 & 30 & 1 & 39 \\
    \textbf{CHASM-SWPC-1407}, $S_1=S_0\setminus M$ & 351 & 365 & 329 & 362 & 1407 \\
    \hline
    \quad CHASM Annotations Flagged Bad, $D$\textsuperscript{*} & 50 & 50 & 65 & 30 & 195 \\
    \quad CHASM Annotations removed by Quality Filter, $Q$ & 105 & 98 & 72 & 32 & 307 \\
    \textbf{CHASM-SWPC-1112}, $S_2=S_1\setminus Q$ & 249 & 267 & 266 & 330 & 1112 \\
    \hline
    \quad SDO Image Sets with $>$1 minute tolerance, $T$ & 53 & 1 & 158 & 2 & 214 \\
    \textbf{CHASM-SWPC-971}, $S_3=S_2\setminus T$ & 217 & 266 & 159 & 329 & 971 \\
    \hline
    Non-unique CHs in CHASM Annotations & 1237 & 1380 & 1038 & 754 & 4409 \\
    Non-unique CHs in SWPC Drawings & 1310 & 1444 & 1046 & 753 & 4553 \\
    \hline
\end{tabular*}
\parbox{\linewidth}{\textit{Note.} An individual CHASM annotation can have multiple CH segmentation masks within it. The exclusion sets ($M$, $Q$, and $T$) overlap, so the drop in dataset size is smaller than the raw count listed for each set. \textsuperscript{*}Set $D$ is a subset of set $Q$.}
\end{table}

\par With a 1 hour matching tolerance, the average (and standard deviation) number of images, over all $7+1=8$ SDO images, was 357.\replaced{3}{0} (1.\replaced{04}{15}) for 2017, 365.0 (0.0) for 2019, 35\replaced{7}{5}.0 (\replaced{7.33}{8.59}) for 2022, and 362.\replaced{88}{78} (0.\replaced{35}{42}) for 2024. Particularly for 2022, there was a substantial number of HMI images that were unavailable.

\par Ultimately, when including all days of those four years in which all 8 SDO images could be matched, we obtained 1\added{,}407 full training input sets. We refer to this dataset as \emph{CHASM-SWPC-1407}.

\par We also performed a further post-processing step by removing every \replaced{CHASM annotation}{synoptic map} containing at least one \replaced{CH}{annotated} segmentation mask that was marked by a labeler as ``bad''. The resulting dataset of 1,11\replaced{2}{1} \replaced{annotations}{images} generally has a better representation of the true CH boundaries than CHASM-SWPC-1407 and is referred to as \emph{CHASM-SWPC-111\replaced{2}{1}}.

Finally, we \replaced{used}{also tried using} a stricter tolerance (1 minute) between the times of the \replaced{SWPC images}{synoptic maps} and satellite images; this resulted in the \emph{CHASM-SWPC-9\replaced{71}{67}} dataset.\added{ All counts and set notation for the datasets can be found in Table \ref{tab:DatasetCounts}.}

In summary, we created the following datasets:

\begin{itemize}
    \item \textbf{CHASM-SWPC-1407}: \deleted{All }1,407 \replaced{CHASM annotations}{days} with complete SDO imagery (maximum data quantity)
    \item \textbf{CHASM-SWPC-111\replaced{2}{1}}: Excludes \replaced{307}{296} \replaced{CHASM annotation-SDO imagery pairs}{days} where any CH\added{ segmentation mask} was flagged as ``bad''\added{ or not all CHs in an SWPC drawing were segmented by SAM} (balanced quality/quantity)
    \item \textbf{CHASM-SWPC-9\replaced{71}{67}}: Excludes \replaced{an additional 214 CHASM annotation-SDO imagery pairs}{440 days} where \deleted{any CH was flagged as "bad" or }any corresponding \replaced{SDO}{AIA} image was outside of 1 minute of the reported time of the SWPC drawing\deleted{.} (maximum quality filtering)
\end{itemize}

\subsection{Dataset Statistics}

\par  The CHASM-SWPC-111\replaced{2}{1} dataset has an average of 3.01\replaced{6}{3} CHs per day. This number varies significantly across years, with fewer detected CHs  around the solar maximum and more around the solar minimum. This trend in our dataset labels agrees with known behaviors in the solar cycle \cite{cranmer_coronal_2009}. 
Specifically, our dataset contains 123\replaced{7}{9} CH \added{segmentation }masks in 2017, 138\replaced{0}{2} CH \added{segmentation }masks in 2019, 1038 CH \added{segmentation }masks in 2022, and 75\replaced{4}{2} CH \added{segmentation }masks in 2024. Note that these are individual CH \emph{\added{segmentation }masks} produced by SAM and selected using CHASM, not the count of unique CHs that year.

\subsection{\replaced{Dataset Biases}{Inter-Annotator Agreement}}
\label{sec:dataset_biases}

\added{Since this dataset includes CH segmentation masks flagged as ``bad'', therefore, some CHASM annotations are not included in the cleaned datasets. The underlying patterns that cause CHASM annotations to be flagged as ``bad'' could potentially create bias in the datasets. In particular, one possible bias is that SWPC drawings with more CHs are more likely to be flagged ``bad'' and not used in the dataset. If any single CH has a poor or missing segmentation boundary, the entire annotation is considered ``bad'', and more CHs in a particular SWPC drawing increases the likelihood that that day will be removed for our clean datasets. Since the annotators manually listed the number of CHs for a given drawing, the number of CHs will be accurate for analysis even in the case CHASM was not able to segment every CH.}

\begin{table}[h!]
    \caption{Correlation Between Number of Coronal Holes and Exclusion from the Cleaned Dataset}
    \label{tab:PearsonCorrelationByYear}
    \centering
    \small
    \begin{tabular*}{\linewidth}{@{\extracolsep{\fill}} l c c c}
        \hline\hline
        Years & Pearson $R$ & $p$-value & Mean Difference (CHs) \\
        \hline
        All Years & 0.088382 & 0.000874 & $+0.3219 \pm 0.0961$ \\
        2017 & -0.090183 & 0.087962 & $-0.2919 \pm 0.1826$ \\
        2019 & 0.100310 & 0.055534 & $+0.3171 \pm 0.1931$ \\
        2022 & 0.137508 & 0.009587 & $+0.4223 \pm 0.1518$ \\
        2024 & 0.190035 & 0.000452 & $+0.6973 \pm 0.2155$ \\
        \hline
    \end{tabular*}
    \parbox{\linewidth}{\textit{Note.} Pearson correlation coefficient ($R$) between the number of coronal holes in an SWPC drawing and that drawing's exclusion from the cleaned dataset, with two-sided $p$-values, and the mean difference in coronal-hole count between excluded and retained drawings ($\pm$ standard error).}
\end{table}


\added{As referenced in Table~\ref{tab:PearsonCorrelationByYear}, the CHASM-SWPC-1112 dataset has a statistically significant bias (with a p-value threshold of 0.05) against SWPC drawings with more CHs over all of the years. However, the behavior between years varies significantly. Only in years 2022 and 2024 is positive correlation between the number of CHs and the likelihood of an annotation to be excluded from our dataset confirmed ($p < 0.05$), while 2017 indicates a potential small reverse bias.}

\added{Another bias to examine within the dataset is between the area of an individual CH compared to its likelihood to be classified as a ``bad'' segmentation. Performing a similar analysis finds that our dataset has a bias against larger CHs, as seen in Table \ref{tab:AreaExclusionCorrelation}, with a correlation between the area of a CH and it being flagged as a ``bad'' segmentation mask. This area was calculated in terms of the number of pixels in a segmentation mask for a particular CH. Because perfect segmentations were not always possible, the areas for ``bad'' CHs are approximations from the annotators for the best possible mask to represent that CH. For each individual CH, the CHASM tool biases slightly against larger CHs, with SAM less likely to create correct segmentation masks for those CHs. This result has high statistical significance, with a weak to moderate correlation. An outlier in this analysis is 2024, which is biased toward larger CHs instead of against them. This may be because 2024 is at a solar maximum, with more CHs of smaller area. We report the distributions of number of coronal holes and coronal hole area before and after removal of bad days as well as statistics about their distributions before and after removal in \ref{sec:appendix_yearly}.}

\begin{table}[h!]
    \caption{Correlation Between Coronal Hole Area and Exclusion from the Cleaned Dataset}
    \label{tab:AreaExclusionCorrelation}
    \centering
    \small
    \begin{tabular*}{\linewidth}{@{\extracolsep{\fill}} l c c c}
        \hline\hline
        Years & Pearson $R$ & $p$-value & Mean Difference (\% disk) \\
        \hline
        All Years & 0.156423 & $<0.000001$ & $+0.6573 \pm 0.1190$ \\
        2017 & 0.323283 & $<0.000001$ & $+2.0530 \pm 0.3254$ \\
        2019 & 0.190779 & $<0.000001$ & $+0.8127 \pm 0.2245$ \\
        2022 & 0.097682 & 0.001628 & $+0.2110 \pm 0.0885$ \\
        2024 & -0.098367 & 0.006943 & $-0.1924 \pm 0.0591$ \\
        \hline
    \end{tabular*}
    \parbox{\linewidth}{\textit{Note.} Pearson correlation coefficient ($R$) between coronal hole area (percent of the solar disk covered) and exclusion from the cleaned dataset, with two-sided $p$-values, and the mean area difference between excluded and retained coronal holes ($\pm$ standard error).}
\end{table}


\textbf{Inter Annotator Agreement}

\par The two annotators of CHASM-SWPC each labeled two years' worth of data: 2017 and 2019 (labeler 1) and 2022 and 2024 (labeler 2). 
To measure the inter-annotator agreement, each annotator additionally labeled 40 random \replaced{SWPC drawings}{days} from each of the two years labeled by the other annotator, without knowledge of the labels the other annotator produced. \added{These overlapping annotations were collected purely for the purposes of inter annotator agreement analysis.} Therefore, a \replaced{160 drawing}{total of 160 days of} overlap \replaced{was}{were} examined, with 40 \replaced{drawings}{days} in each of the years\added{:} 2017, 2019, 2022, and 2024. 

\par Over these days, we found a mean intersection-over-union (IoU) between annotators of 0.975 on CHASM-SWPC-1407 and 0.988 on CHASM-SWPC-111\replaced{2}{1}. \added{This IoU represents a pixel wise calculation on the full disk images of the masks with all CHs merged.}

\par
\begin{figure}[h!]
    \centering
    \includegraphics[width=0.6\linewidth]{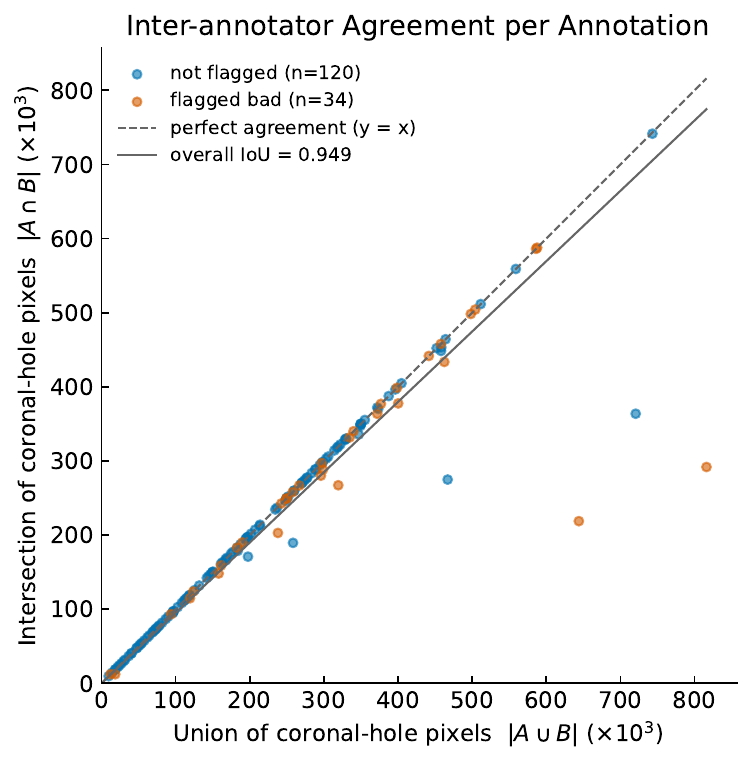}
    \caption{Intersection and union between annotators for the 160 days with shared annotations.}
    \label{fig:anno_iou}
\end{figure}

 Furthermore, the two annotators agreed on the flagging of which \replaced{annotations}{days} were ``bad'' for 86.88\% of the 160 \replaced{drawings}{days}. The \added{SWPC} CH confidence values input by the two labelers agreed with each other \replaced{in}{on} 158 (98.75\%) of 160 \replaced{annotations}{days}. \added{Figure~\ref{fig:anno_iou} shows the intersection and union for the 160 annotations used to perform this analysis. We find that nearly all days match exactly between annotators, with outliers that required subjective judgments, where no single option using the tool optimally captured the image, and each annotator tried to best replicate the image in diverging ways. }

\added{There are a few outliers that sit to the far right of the graph and far from the ideal (intersection=union) perfect agreement line. For outliers in this region that are flagged correctly as bad, we find the annotators attempted differing workarounds to undesirable SAM masks. The blue labels (that have large inter-annotator disagreement despite not being flagged) are human error from the annotators, and represent a small minority of the dataset. These points represent approximately 2-4 annotations over this sample of 160, or 1.25\%-2.5\% of labels. Extrapolating this to the full dataset, we can estimate coarsely somewhere between 18 and 36 mislabeled annotations that are inappropriately not flagged bad across the 1446 annotations.}

\subsection{Qualitative Comparison to SPoCA-CH Labels}
\begin{figure*}[!htbp]
    \centering
    \includegraphics[width=.95\textwidth]{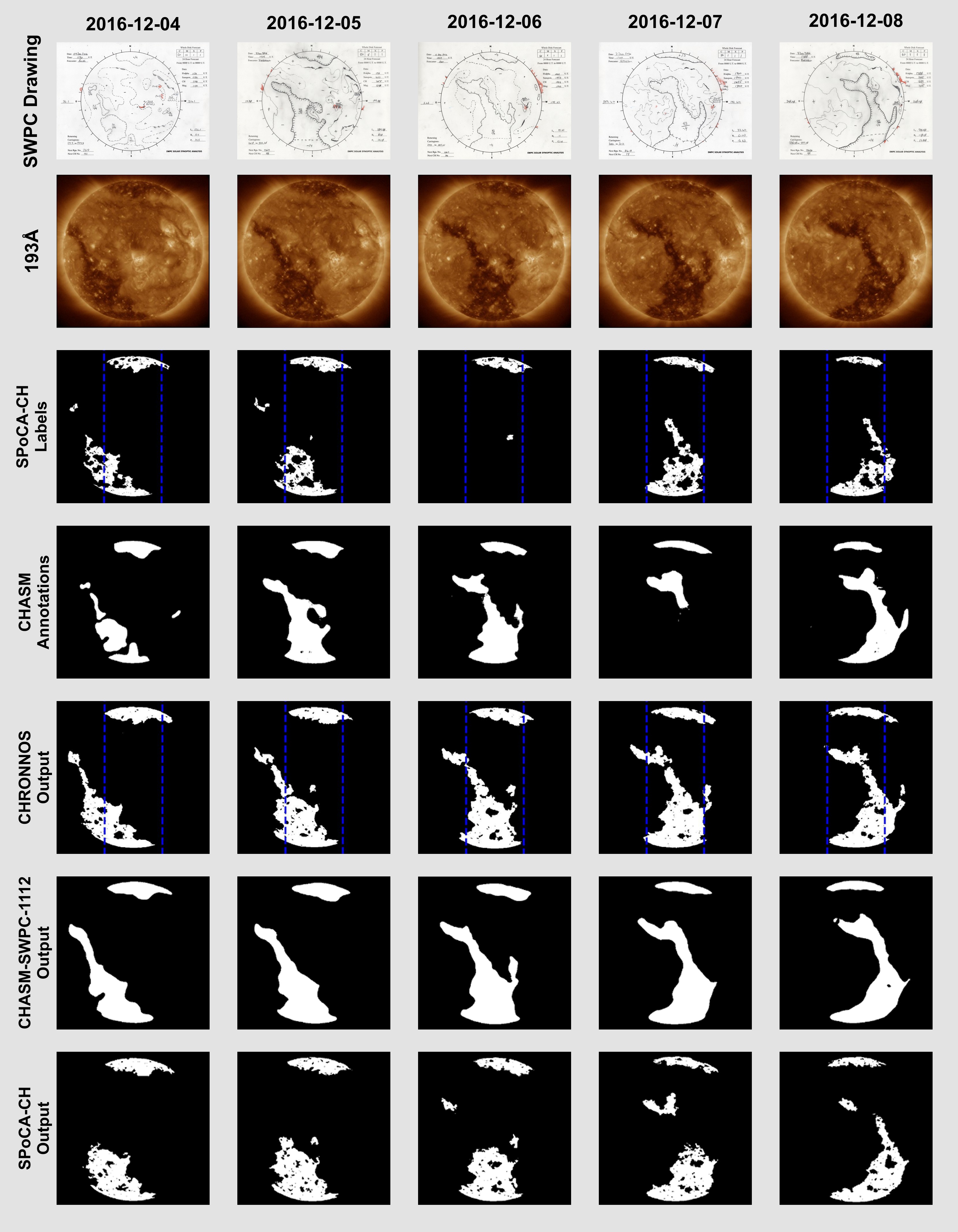}
    \caption{Comparison of CH segmentations across multiple methods from 2016-12-04 to 2016-12-08. Rows show: SWPC \deleted{synoptic }drawings (row 1), corresponding 193\AA\space imagery (row 2), SPoCA-CH labels (row 3), CHASM \replaced{annotations}{-SWPC-1111 labels} (row 4), CHRONNOS predictions \added{from \protect\cite{jarolim_multi-channel_2021}} (row 5), CHASM-SWPC-111\replaced{2}{1} model predictions (row 6), and \replaced{predictions from the publicly available CHRONNOS model}{pre-trained CHRONNOS predictions trained on SPoCA-CH} (row 7). \replaced{The CHASM annotations demonstrated here are purely for qualitative comparison with the existing labels for SPoCA-CH. With our annotation tool, the 2016-12-07 sample was flagged as “bad.”}{The 2016-12-07 sample was flagged as ``bad'' during CHASM annotation and excluded from CHASM model training.} Credit: Jarolim et al., A\&A, 652, A13, 2021, reproduced with permission © ESO.}
    \label{fig:5days}
\end{figure*}

\citeA{jarolim_multi-channel_2021} trained the CHRONNOS neural network on a dataset that was labeled automatically by the SPoCA algorithm and then adjusted by manual annotation of CHs versus solar filaments \cite{verbeeck_spoca-suite_2013}; the resulting label set is called SPoCA-CH. In contrast, the labels of CHASM-SWPC are derived from expert human annotation, followed by algorithmic extraction (SAM), and then refined again by manual annotation using our CHASM tool. Figure \ref{fig:5days} (rows 3 and 4) gives a qualitative comparison of \replaced{annotations created with CHASM}{labels from CHASM-SWPC-1111} to SPoCA-CH \added{for days in 2016 which are not included in the CHASM-SWPC dataset}. Compared to the SWPC \replaced{drawings}{synoptic maps} (row 1) and 193\AA\space imagery (row 2), \replaced{CHASM annotations tend}{CHASM-SWPC-1111 tends} to cover the CH area more completely than the SPoCA-CH labels. An exception is 2016-12-07 (column 4), on which \added{day} \replaced{the CHASM annotation}{CHASM-SWPC-1111} is missing a significant chunk of a CH; however, due to the manual annotation step used in CHASM, this day was appropriately flagged as ``bad''\deleted{ and thus not used for training}. \added{On these held out days, we find the model is able to generalize beyond the capabilities of the annotation pipeline; on 2016-12-07 the model is able to outperform the segmentation of the annotation pipeline itself.} Finally, we observe that CHASM-SWPC is spatially smoother, whereas SPoCA-CH often exhibits a ``Swiss cheese'' (many small holes within the CHs) effect. \added{Characteristics of each dataset carry through training, and the characteristics of SPoCA-CH may appropriately represent CH boundaries under an annotation approach grounded in intensity thresholding and clustering. CHASM-SWPC helps train models which reproduce SWPC-style expert annotations.}

\subsection{Dataset Biases}
\label{DatasetBiases}

\par Between CHASM-SWPC-1407 and CHASM-SWPC-111\replaced{2}{1}, we removed all days in which at least one CH was flagged by an annotator as ``bad''. This yields a dataset in which the CH boundaries are more accurate, but it also introduces the possibility of biases dependent on the probability of a ``bad'' segmentation mask. For downstream analyses based on CHASM-SWPC labels -- e.g., an investigation into the relationship between CH size and number at different parts of the solar cycle -- these biases could be important to take into account.
We thus assessed the possibility of (1) {\em count bias}: whether a correlation existed between flagging a day as ``bad'' and the number of CHs on that day; and (2) {\em size bias}: whether a correlation existed between flagging a day as ``bad'' and the size (number of pixels according to the CHASM-based labels) of the CHs on that day.

\par {\em Count bias}: The correlations between the number of CHs on each day and flagging that day as ``bad'' were  $-0.090$ ($p=0.088$), $0.100$ ($p=0.056$), $0.147$ ($p=0.005$) and $0.204$ ($p<0.001$) for 2017, 2019, 2022, and 2024, respectively. Over all four years, the correlation was $0.103$ ($p<0.001$). 

\par {\em Size bias}: The correlations between the average number of pixels  assigned to a CH on each day and flagging that day as ``bad'' were $0.324$ ($p<0.001$), $0.188$ ($p<0.001$), $0.051$ ($p=0.102$), and $-0.104$ ($p=0.004$). Over all four years, the correlation was $0.146$ ($p<0.001$). The negative  correlation in 2024 could possibly be due to CH behavior near the solar maximum.

\par These results suggest that SAM's segmentations are slightly less accurate on days with larger CHs (size bias) and on days with more CHs (count bias). These  effects can change over the solar cycle, such as the correlation \replaced{of}{from} size bias changing from positive to negative across the years from solar minimum (2017) to solar maximum (2024). While these biases are small, they are important to keep in mind, especially for future researchers who plan to analyze CH behavior across the solar cycle.


\section{Experiments Training on CHASM-SWPC}

To investigate how \added{useful} the CHASM-SWPC datasets \replaced{are}{can be useful} for training CH detection models, we compared a CHRONNOS neural network trained from scratch using CHASM-SWPC-1407, CHASM-SWPC-111\replaced{2}{1}, or CHASM-SWPC-9\replaced{71}{67}, to the same CHRONNOS network trained on SPoCA-CH pseudo-labels (like in \citeA{jarolim_multi-channel_2021}). This comparison serves two purposes: (1) validating that expert-derived ground truth (CHASM-SWPC) improves model performance compared to algorithmically-generated pseudo-labels (SPoCA-CH), and (2) investigating the tradeoff between dataset size and annotation quality across the three CHASM-SWPC variants. 

In all cases, we evaluate the trained model on the CHASM-SWPC-111\replaced{2}{1} test set, consisting of all days in November and December from years 2017, 2019, 2022, and 2024. \added{All days in November and December were excluded from training all of our CHASM-SWPC model variants, reserving them exclusively for testing each variant. The original SPoCA-CH-trained CHRONNOS model was not trained on the same calendar years, and the months of November and December were used as the test set during its training procedure as well. Due to the temporal nature of solar data, utilizing a calendar-based split allows for longer timespan forecasting since the test set samples are not similar to any of the training samples, as opposed to randomly selecting days throughout a calendar year as a test set. Doing so may lead to the memorization of similar observations during training, as explained in \protect\citeA{jarolim_multi-channel_2021}.} We also qualitatively (visually) compared the outputs of CHRONNOS when trained on CHASM-SWPC-111\replaced{2}{1} to the same model trained on SPoCA-CH\added{, as seen in Figure \ref{fig:5days}}.\deleted{Under the assumption that the SWPC synoptic maps, which are hand-annotated by experts at NOAA, are more faithful indicators of the CHs’ true positions compared to pseudo-labels obtained from the SPoCA-CH algorithm, then higher accuracy on the CHASM-SWPC-1111 test set would imply a more accurate model.}

Consistent with prior work 
\cite{jarolim_multi-channel_2021, wang_solar_2025, TSS_FlareForecasting_Barnes}, we evaluate models using three different metrics:

\begin{enumerate}
    \item \textbf{Accuracy}: Fraction of all pixels correctly classified.
    \item \textbf{True Skill Statistic (TSS) \cite{bloomfield2012toward}}, calculated as TSS = Sensitivity  + Specificity   - 1 
    \item \textbf{Intersection over Union (IoU)} between the detected and ground truth CH regions.
\end{enumerate}

Since CHRONNOS outputs probabilistic predictions, we apply a threshold $\theta \in [0,1]$ to produce binary classifications, where pixels with probability $> \theta$ are classified as CHs. For each model, we report results at two threshold settings: (1) $\theta = 0.5$ (non-optimized baseline), and (2) an optimized $\theta$ selected to maximize the average of Accuracy, TSS, and IoU\deleted{ on the training set}\added{for each model} (see \ref{sec:appendix_thresholding} for optimal thresholds per model). \added{Threshold optimization was performed exclusively on each model's training set. All reported threshold optimization metrics were then computed on the model's test set predictions using this training-derived threshold. Following \protect\citeA{jarolim_multi-channel_2021}, we did not have any validation set throughout our experimentation, therefore, threshold optimization was done entirely on each models' training set.}

All models were evaluated on the CHASM-SWPC-111\replaced{2}{1} test set, thus ensuring consistent comparison across training datasets. Following the CHRONNOS testing methodology \cite{jarolim_multi-channel_2021}, we report metrics \replaced{for both}{both for} the full solar disk and for the central band between $-400''$ and $+400''$ solar longitude. All models were trained according to the training scheme outlined by \citeA{jarolim_multi-channel_2021} for CHRONNOS, including the same hyperparameters \deleted{as }used for training CHRONNOS on SPoCA-CH (which removed the need for a separate validation set for hyperparameter optimization). \added{All training parameters can be found in \protect\ref{sec:appendix_training_details}.} 

\subsection{Comparison of CHASM-SWPC to SPoCA-CH} 
\label{sec:CHASM_vs_SPoCA}
The full disk results (Table~\ref{tab:FullDiskMainTable}) and central band results (Table~\ref{tab:CentralBandMainTable}) demonstrate that models trained on CHASM-SWPC datasets \deleted{substantially }outperform the model trained on SPoCA-CH when evaluated against expert annotations. \added{Against the CHASM-SWPC-1112 test set,} at optimized thresholds, the CHASM-SWPC-1112-trained model achieves \added{an} IoU of 0.5668 on the full disk, representing a 1\replaced{8.0}{7.8}\% relative improvement over the SPoCA-CH-trained model (IoU = 0.4805). \added{CHASM-SWPC-1112's model accuracy and TSS had a small $\sim$1.0\% increase over SPoCA-CH for both models at optimized thresholds.}\deleted{Similarly, TSS improves from 0.6749 to 0.\replaced{6807}{7811} \replaced{, a small 0.9\% improvement}{(15.7\% relative improvement)}. } \added{We do note that not all metrics increased with optimized thresholds, in fact TSS decreased for all CHASM-SWPC model metrics. Accuracy and IoU improved when thresholding the CHASM-SWPC models. The opposite is true for the SPoCA-CH-trained model: accuracy and IoU decreased slightly, while TSS improved.} These gains demonstrate that training on expert-derived ground truth produces segmentations more consistent with expert consensus than training on algorithmically-generated \deleted{pseudo-}labels.

\begin{table}[h!]
    \caption{Full Disk Results}
    \label{tab:FullDiskMainTable}
    \centering
    \small
    \begin{tabular*}{\linewidth}{@{\extracolsep{\fill}} l c c c c}
        \hline\hline
        Training Set & Threshold & Accuracy & TSS & IoU \\
        \hline
        CHASM-SWPC-1407 & 0.5000 & 0.9687 & \underline{0.7597} & 0.4985 \\
        CHASM-SWPC-1407 & 0.9308 & \underline{0.9793} & 0.6256 & 0.5316  \\
        CHASM-SWPC-111\replaced{2}{1} & 0.5000 & 0.9740 & \textbf{0.7811} & \underline{0.5413} \\
        CHASM-SWPC-111\replaced{2}{1} & 0.9328 &\textbf{0.9802} & 0.6807 & \textbf{0.5668} \\
        CHASM-SWPC-9\replaced{71}{67} & 0.5000 & 0.9718 & 0.7633 & 0.5227 \\
        CHASM-SWPC-9\replaced{71}{67} & 0.8976 & 0.9781 & 0.6875 & 0.5495 \\
        SPoCA-CH & 0.5000 & 0.9759 & 0.5626 & 0.4813 \\
        SPoCA-CH & 0.0039 & 0.9708 & 0.6749 & 0.4805 \\
        \hline
    \end{tabular*}
    \parbox{\linewidth}{\textit{Note.} Metrics of models based on training dataset evaluated on full solar-disk of CHASM-SWPC-111\replaced{2}{1} test set. Thresholds of value 0.5 are unoptimized, with other threshold values selected to maximize the average of Accuracy, IoU, and TSS metrics. \textbf{Bold} is the highest and \underline{underlined} is the second highest value in the column.}
\end{table}

\begin{table}[h!]
    \caption{Central Band Results}
    \label{tab:CentralBandMainTable}
    \centering
    \small
    \begin{tabular*}{\linewidth}{@{\extracolsep{\fill}} l c c c c}
        \hline\hline
        Training Set & Threshold & Accuracy & TSS & IoU \\
        \hline
        CHASM-SWPC-1407 & 0.5000 & 0.9459 & 0.7286 & 0.5026 \\
        CHASM-SWPC-1407 & 0.9355 & 0.9651 & 0.6159 & 0.5348 \\
        CHASM-SWPC-111\replaced{2}{1} & 0.5000 & 0.9559 & \textbf{0.7604} & \underline{0.5453}  \\
        CHASM-SWPC-111\replaced{2}{1} & 0.9338 & \textbf{0.9667} & 0.6769 & \textbf{0.5707} \\
        CHASM-SWPC-9\replaced{71}{67} & 0.5000 & 0.9524 & \underline{0.7414} & 0.5263 \\
        CHASM-SWPC-9\replaced{71}{67} & 0.9000 & \underline{0.9633} & 0.6804 & 0.5552 \\
        SPoCA-CH & 0.5000 & 0.9592 & 0.5868 & 0.5007 \\
        SPoCA-CH & 0.0112 & 0.9543 & 0.6688 & 0.5069 \\
        \hline
    \end{tabular*}
    \parbox{\linewidth}{\textit{Note.} Metrics of models based on training dataset evaluated on central solar-disk band \([-400'',400'']\) \added{of} CHASM-SWPC-111\replaced{2}{1} test set. Thresholds of value 0.5 are unoptimized, with other threshold values selected to maximize the average of Accuracy, IoU, and TSS metrics. \textbf{Bold} is the highest and \underline{underlined} is the second highest value in the column.}
\end{table}

\begin{figure}[h]
    \centering
    \includegraphics[width=0.9\linewidth]{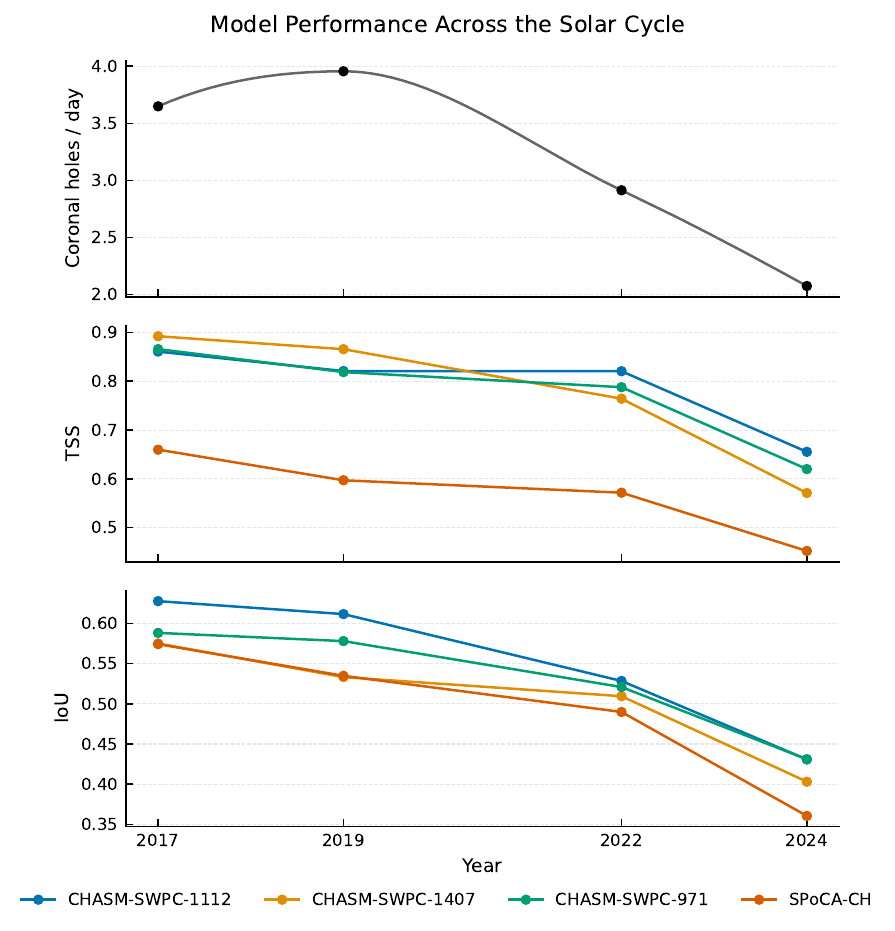}
    \caption{The top figure illustrates the average number of coronal holes per day across the validation set of CHASM-SWPC-1112. The bottom two figures show the results of the models trained on CHASM-SWPC datasets as well as the original CHRONNOS model trained on the SPoCA-CH dataset across various testing years throughout the solar cycle. These models are each evaluated at a threshold of 0.5.}
    \label{fig:across_years}
\end{figure}

\added{We additionally compare the results across these metrics for each model across each validation year, and find there are trends in performance across the solar cycle. We find, across the solar cycle, all models have similar trends in performance: higher performance correlating with more coronal holes per day. We believe this may be due to the fixed thresholding of these models in Figure \ref{fig:across_years}, and believe thresholding of these models depending on the period of the solar cycle may yield higher performance than any singular metric. See \protect\ref{sec:appendix_yearly} for detailed tables of these results across the years of the solar cycle.}

\subsection{Comparison of CHASM-SWPC Datasets Variants}

Training CHRONNOS on three progressively filtered CHASM-SWPC variants reveals how annotation quality affects model performance. CHASM-SWPC-111\replaced{2}{1} achieves the best results across most metrics, validating our hypothesis that it achieves a balance between data quality and quantity. With 29\replaced{5}{6} fewer \replaced{data samples}{training days} than CHASM-SWPC-1407, the CHASM-SWPC-111\replaced{2}{1}-trained model achieves 6.6\% higher IoU on the full disk (0.5668 vs. 0.5316 at optimized thresholds). CHASM-SWPC-1407's slightly lower performance likely results from including days with \replaced{segmentations flagged as ``bad''}{inaccurate segmentations}. Conversely, CHASM-SWPC-9\replaced{71}{67} achieves slightly worse performance (except for TSS with an optimized threshold) than CHASM-SWPC-111\replaced{2}{1}, suggesting that further filtering provides diminishing returns.

\deleted{Figure~\ref{fig:5days} shows qualitative comparisons for five days in the test set where both SPoCA-CH and CHASM-SWPC-1111 ground truth labels are available. SPoCA-CH labels exhibit a characteristic ``Swiss cheese'' pattern with numerous small internal voids (row 3), and the model trained on SPoCA-CH reproduces this fragmented appearance (row 5), thus failing to generalize to the smooth, contiguous boundaries seen in expert SWPC drawings (row 1). In contrast, CHASM-SWPC-1111 labels (row 4) feature spatially smooth boundaries without internal holes, and the CHASM-SWPC-1112 model trained on this dataset produces similar smooth outputs (row 6) that closely match the SWPC drawings (row 1). A notable example is 2016-12-07 (column 4), where CHASM's semi-automated extraction failed to capture the full CH extent visible in the SWPC drawing. This day was appropriately flagged as ``bad'' by annotators and thus excluded from CHASM-SWPC-1111 training. Moreover, the neural network trained using CHASM-SWPC-1111 successfully predicted the CH boundaries for that day (even though the SAM masks themselves were inaccurate), thus demonstrating strong generalization capability.}

\section{Discussion}

Our experiments demonstrate three key findings with implications for automated CH detection using machine learning:

\replaced{Training on CHASM-SWPC data substantially increases the model performance by 18.0\%, when evaluated on CHASM-SWPC test data, compared to training the exact same model on algorithmically-generated pseudo-labels (SPoCA-CH data). The difference in the models' predictions is particularly pronounced for full-disk analysis where previous methods struggled.}{Training on expert-derived ground truth substantially improves model performance, compared to training on algorithmically-generated pseudo-labels, when the expert-derived test set is also used for evaluation. The 18.0\% improvement in IoU represents a meaningful advance in CH detection capability, particularly for full-disk analysis where previous methods struggled.Training on expert-derived ground truth provides a different quantitative and qualitative view of coronal holes. The 18.0\% improvement between SPoCA-CH trained models and CHASM trained models demonstrates this difference.} Assuming the expert\deleted{-drawn} SWPC \replaced{drawings}{synoptic maps} are more faithful representations of the true CH positions than SPoCA-CH \deleted{pseudo-}labels, this result demonstrates that the source of  training set labels has a substantial impact. It is possible that digitizing historical expert annotations for other solar features (e.g., active regions, filaments) could also be beneficial.

2) Curating the dataset by systematically removing days with poor segmentations can yield\deleted{s} better models. CHASM-SWPC-111\replaced{2}{1}, with 21\% fewer training days than CHASM-SWPC-1407, achieves superior performance across all metrics. Researchers using CHASM datasets should prioritize CHASM-SWPC-111\replaced{2}{1} for model training, while CHASM-SWPC-1407 may be more suitable for statistical analyses requiring maximum temporal coverage.

3) CH detection models trained on high-quality, curated expert data generalize well to unseen data\deleted{,} and can even be more accurate than the ground truth labels if the annotation process itself failed (e.g., 2016-12-07 as shown in Figure \ref{fig:5days}). This demonstrates that systematic removal of identified low-quality examples does not harm generalization, complementing prior observations that models can tolerate sparse ground truth errors \cite{jarolim_multi-channel_2021}. Our results suggest that when quality assessment is feasible -- as enabled by CHASM's manual flagging -- active curation is preferable to passive tolerance of noise.

\textbf{Limitations}: As documented in Section~\ref{DatasetBiases}, quality filtering introduces modest statistical biases toward smaller CHs and against days with many CHs. While these biases do not appear to harm model performance for detection tasks, researchers using CHASM-SWPC datasets for scientific analyses of CH size distributions or solar cycle dependencies should account for these selection effects. \added{Also, we note that SWPC annotations, while performed by experts, are still hand-drawn and not direct pixel-level physical measurements.}

\textbf{Future work}:
Testing on other neural network architectures (e.g., transformer-based models \cite{wang_solar_2025}) would further validate the generalization of our findings.

\section{Conclusions}

\par We have developed CHASM, a semi-automated pipeline to digitize solar observations previously inaccessible to automated detection algorithms. Leveraging the Segment Anything Model \cite{kirillov_segment_2023}, CHASM enabled \replaced{human labelers}{us} to \replaced{annotate}{process} 1,446 NOAA SWPC \replaced{drawings}{synoptic maps} in approximately 14 hours\deleted{ of human annotation time}, producing accurate segmentations for over 4,4\replaced{09}{11} \added{non-unique} coronal holes. Agreement between segmentations made by different annotators is high, with an IoU of 0.988 on the CHASM-SWPC-111\replaced{2}{1} dataset, indicating \added{that }the CHASM tool allows different users to create very similar segmentation masks with minimal error. 

\par CHASM's additional features, including manual flagging, CH counting, \added{and capture of SWPC-assigned, per CH} confidence \deleted{scoring, }and polarity\deleted{labeling,} enable systematic dataset refinement and support diverse downstream analyses. We publicly release all CHASM-SWPC datasets, the annotation tool, and trained CHRONNOS model weights for use in analyzing similar solar observations\added{ in the future}, such as \replaced{solar disk drawings}{synoptic maps} from the Met Office Space Weather Operations Centre (MOSWOC) in the United Kingdom. We release three dataset variants: CHASM-SWPC-1407 (all days with complete imagery), CHASM-SWPC-111\replaced{2}{1} (excluding days with flagged poor segmentations), and CHASM-SWPC-9\replaced{71}{67} (maximum quality filtering), allowing others to investigate quality-quantity tradeoffs in their own applications. 

\par Training CHRONNOS \cite{jarolim_multi-channel_2021} on the CHASM-SWPC-111\replaced{2}{1} dataset achieved an accuracy of 0.9802, a TSS of 0.\replaced{6807}{7811}, and an IoU of 0.5668 when optimizing the threshold. Compared to the original CHRONNOS model trained on SPoCA-CH at an optimal threshold, this represents an 1\replaced{8.0}{7.8}\% relative improvement in IoU when evaluated on the CHASM-SWPC-111\replaced{2}{1} test set. We release the CHRONNOS model trained on the CHASM-SWPC-111\replaced{2}{1} dataset, as well as optimal thresholds for the model and its metrics (\ref{sec:appendix_thresholding}).

\appendix
\section{Model Results by Year}
\label{sec:appendix_yearly}

\par Due to the varying distribution, size, and number of coronal holes throughout the solar cycle, we report our results across the test sets of each year for our CHASM-SWPC-1112 dataset in Table~\ref{tab:FullDiskYearlyTable} and Table~\ref{tab:CentralBandYearlyTable} below. All models are thresholded at 0.5.

\begin{table}[h]
    \caption{Full Disk Results by Year}
    \label{tab:FullDiskYearlyTable}
    \centering
    \small
    \begin{tabular*}{\linewidth}{@{\extracolsep{\fill}} l c c c c c}
        \hline\hline
        Training Set & Year & N & Accuracy & TSS & IoU \\
        \hline
        CHASM-SWPC-1112 & 2017 & 45 & \underline{0.9699} & 0.8609 & \textbf{0.6276} \\
        CHASM-SWPC-1407 & 2017 & 45 & 0.9599 & \textbf{0.8925} & 0.5750 \\
        CHASM-SWPC-971 & 2017 & 45 & 0.9634 & \underline{0.8661} & \underline{0.5881} \\
        SPoCA-CH & 2017 & 45 & \textbf{0.9710} & 0.6599 & 0.5741 \\
        \hline
        CHASM-SWPC-1112 & 2019 & 43 & \underline{0.9695} & \underline{0.8211} & \textbf{0.6115} \\
        CHASM-SWPC-1407 & 2019 & 43 & 0.9545 & \textbf{0.8659} & 0.5332 \\
        CHASM-SWPC-971 & 2019 & 43 & 0.9648 & 0.8189 & \underline{0.5779} \\
        SPoCA-CH & 2019 & 43 & \textbf{0.9696} & 0.5971 & 0.5348 \\
        \hline
        CHASM-SWPC-1112 & 2022 & 47 & 0.9787 & \textbf{0.8207} & \textbf{0.5284} \\
        CHASM-SWPC-1407 & 2022 & 47 & 0.9790 & 0.7644 & 0.5094 \\
        CHASM-SWPC-971 & 2022 & 47 & \underline{0.9791} & \underline{0.7880} & \underline{0.5208} \\
        SPoCA-CH & 2022 & 47 & \textbf{0.9828} & 0.5716 & 0.4898 \\
        \hline
        CHASM-SWPC-1112 & 2024 & 57 & 0.9768 & \textbf{0.6554} & \underline{0.4308} \\
        CHASM-SWPC-1407 & 2024 & 57 & \underline{0.9780} & 0.5710 & 0.4031 \\
        CHASM-SWPC-971 & 2024 & 57 & 0.9778 & \underline{0.6200} & \textbf{0.4310} \\
        SPoCA-CH & 2024 & 57 & \textbf{0.9787} & 0.4522 & 0.3606 \\
        \hline
    \end{tabular*}
    \parbox{\linewidth}{\textit{Note.} Metrics for each model evaluated for the full solar disk on the CHASM-SWPC-1112 test set. All models are evaluated at a threshold of 0.5. Emphasis is applied within each year. N is the number of test days in that year. \textbf{Bold} is the highest and \underline{underlined} is the second highest value in the column for that year.}
\end{table}

Table~\ref{tab:FullDiskYearlyTable} reports the results plotted in Figure~\ref{fig:across_years}. Table~\ref{tab:CentralBandYearlyTable} is not reported graphically.

\newpage
\begin{table}[h]
    \caption{Central Band Results by Year}
    \label{tab:CentralBandYearlyTable}
    \centering
    \small
    \begin{tabular*}{\linewidth}{@{\extracolsep{\fill}} l c c c c c}
        \hline\hline
        Training Set & Year & N & Accuracy & TSS & IoU \\
        \hline
        CHASM-SWPC-1112 & 2017 & 45 & \underline{0.9454} & \underline{0.8599} & \textbf{0.6590} \\
        CHASM-SWPC-1407 & 2017 & 45 & 0.9295 & \textbf{0.8732} & 0.6122 \\
        CHASM-SWPC-971 & 2017 & 45 & 0.9382 & 0.8583 & 0.6337 \\
        SPoCA-CH & 2017 & 45 & \textbf{0.9516} & 0.7389 & \underline{0.6514} \\
        \hline
        CHASM-SWPC-1112 & 2019 & 43 & \underline{0.9413} & \underline{0.8368} & \textbf{0.6619} \\
        CHASM-SWPC-1407 & 2019 & 43 & 0.9117 & \textbf{0.8446} & 0.5830 \\
        CHASM-SWPC-971 & 2019 & 43 & 0.9314 & 0.8246 & \underline{0.6283} \\
        SPoCA-CH & 2019 & 43 & \textbf{0.9429} & 0.6642 & 0.6116 \\
        \hline
        CHASM-SWPC-1112 & 2022 & 47 & 0.9687 & \textbf{0.7454} & \textbf{0.4828} \\
        CHASM-SWPC-1407 & 2022 & 47 & 0.9685 & 0.6868 & 0.4575 \\
        CHASM-SWPC-971 & 2022 & 47 & \underline{0.9694} & \underline{0.7207} & \underline{0.4654} \\
        SPoCA-CH & 2022 & 47 & \textbf{0.9732} & 0.5286 & 0.4296 \\
        \hline
        CHASM-SWPC-1112 & 2024 & 57 & 0.9647 & \textbf{0.6367} & \textbf{0.4190} \\
        CHASM-SWPC-1407 & 2024 & 57 & \underline{0.9661} & 0.5613 & 0.3928 \\
        CHASM-SWPC-971 & 2024 & 57 & 0.9654 & \underline{0.6032} & \underline{0.4146} \\
        SPoCA-CH & 2024 & 57 & \textbf{0.9661} & 0.4565 & 0.3568 \\
        \hline
    \end{tabular*}
    \parbox{\linewidth}{\textit{Note.} Metrics for each model evaluated for the central band [-400", 400"] on the CHASM-SWPC-1112 test set. All models are evaluated at a threshold of 0.5. Emphasis is applied within each year. N is the number of test days in that year. \textbf{Bold} is the highest and \underline{underlined} is the second highest value in the column for that year.}
\end{table}

These results mirror those of Section~\ref{sec:CHASM_vs_SPoCA}, which provides more details. We again report that the results for all metrics vary across the solar cycle, and thresholding depending on the period of the cycle may improve performance.

We report the distributions of CH area in Figure~\ref{fig:area_distribution} and in Table~\ref{tab:AreaDistributionStats} below. We find this aligns with the solar cycle, and exhibits a long right tail. Note that Figure~\ref{fig:area_distribution} varies from the area bias reported in Section~\ref{sec:dataset_biases} which reports area on a per day basis on aggregated CHs.

\begin{figure}[h]
    \centering
    \includegraphics[width=0.75\linewidth]{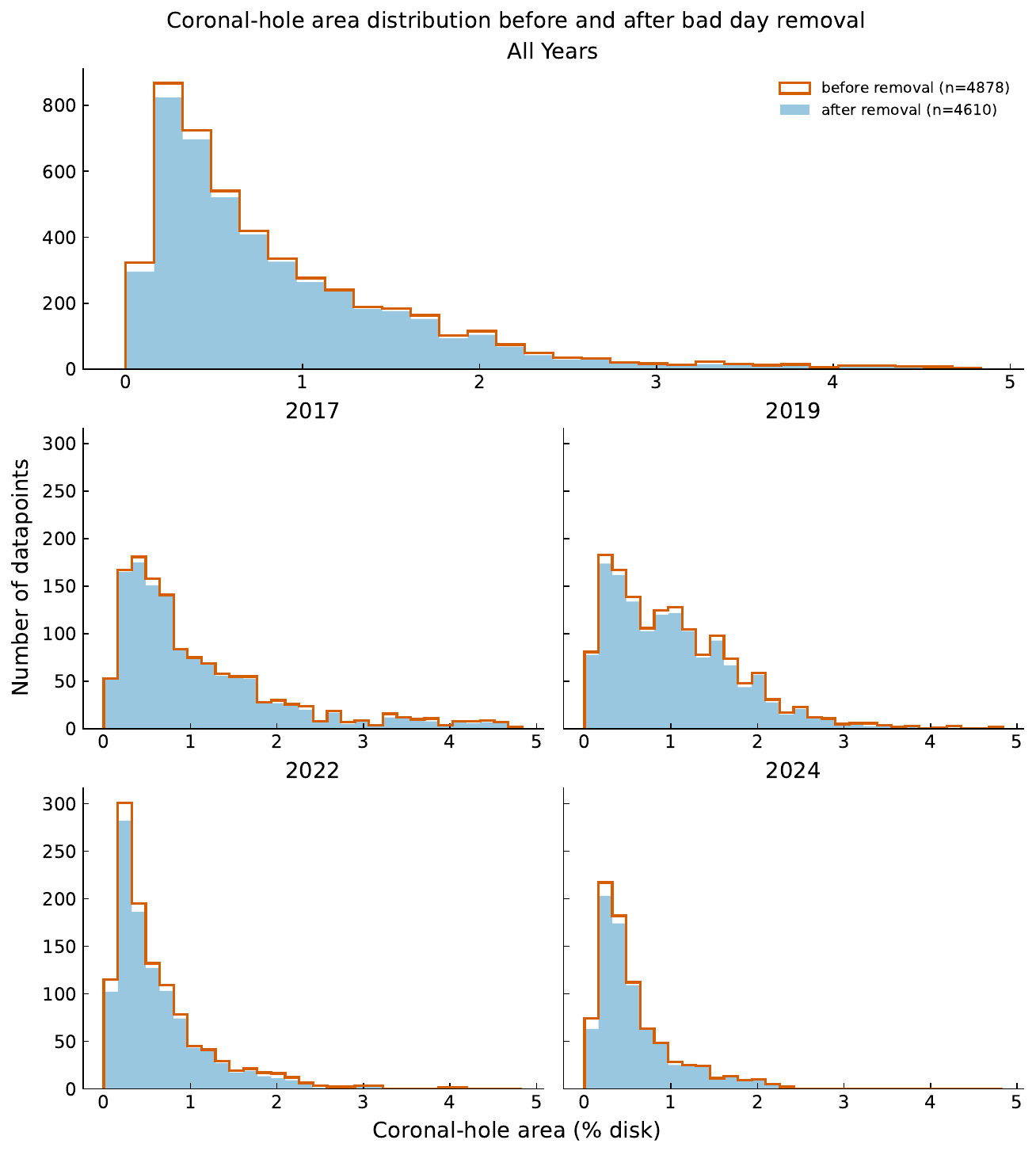}
    \caption{CH area distribution of individual CHs from the dataset before and after removal of bad CHs.}
    \label{fig:area_distribution}
\end{figure}

\begin{table}[H]
    \caption{Coronal Hole Area Distribution Before and After Bad Day Removal}
    \label{tab:AreaDistributionStats}
    \centering
    \small
    \begin{tabular*}{\linewidth}{@{\extracolsep{\fill}} l l c c c c c}
        \hline\hline
        Years & Stage & $N$ & Mean & Median & SD & Skewness \\
        \hline
        All Years & Before & 4878 & 0.939 & 0.635 & 0.948 & 2.956 \\
        All Years & After & 4610 & 0.901 & 0.630 & 0.858 & 2.725 \\
        \hline
        2017 & Before & 1374 & 1.252 & 0.795 & 1.277 & 2.348 \\
        2017 & After & 1310 & 1.162 & 0.774 & 1.136 & 2.339 \\
        \hline
        2019 & Before & 1528 & 1.081 & 0.911 & 0.888 & 2.650 \\
        2019 & After & 1447 & 1.043 & 0.898 & 0.806 & 2.159 \\
        \hline
        2022 & Before & 1153 & 0.644 & 0.457 & 0.590 & 2.485 \\
        2022 & After & 1071 & 0.626 & 0.451 & 0.551 & 2.321 \\
        \hline
        2024 & Before & 823 & 0.568 & 0.426 & 0.445 & 1.621 \\
        2024 & After & 782 & 0.578 & 0.437 & 0.447 & 1.598 \\
        \hline
    \end{tabular*}
    \parbox{\linewidth}{\textit{Note.} Summary statistics of CH area (percent of the solar disk) for the full set (before) and the set retained after bad day removal (after), by year and pooled. SD is the sample standard deviation and skewness is the adjusted Fisher-Pearson coefficient.}
\end{table}

\begin{figure}[h]
    \centering
    \includegraphics[width=0.75\linewidth]{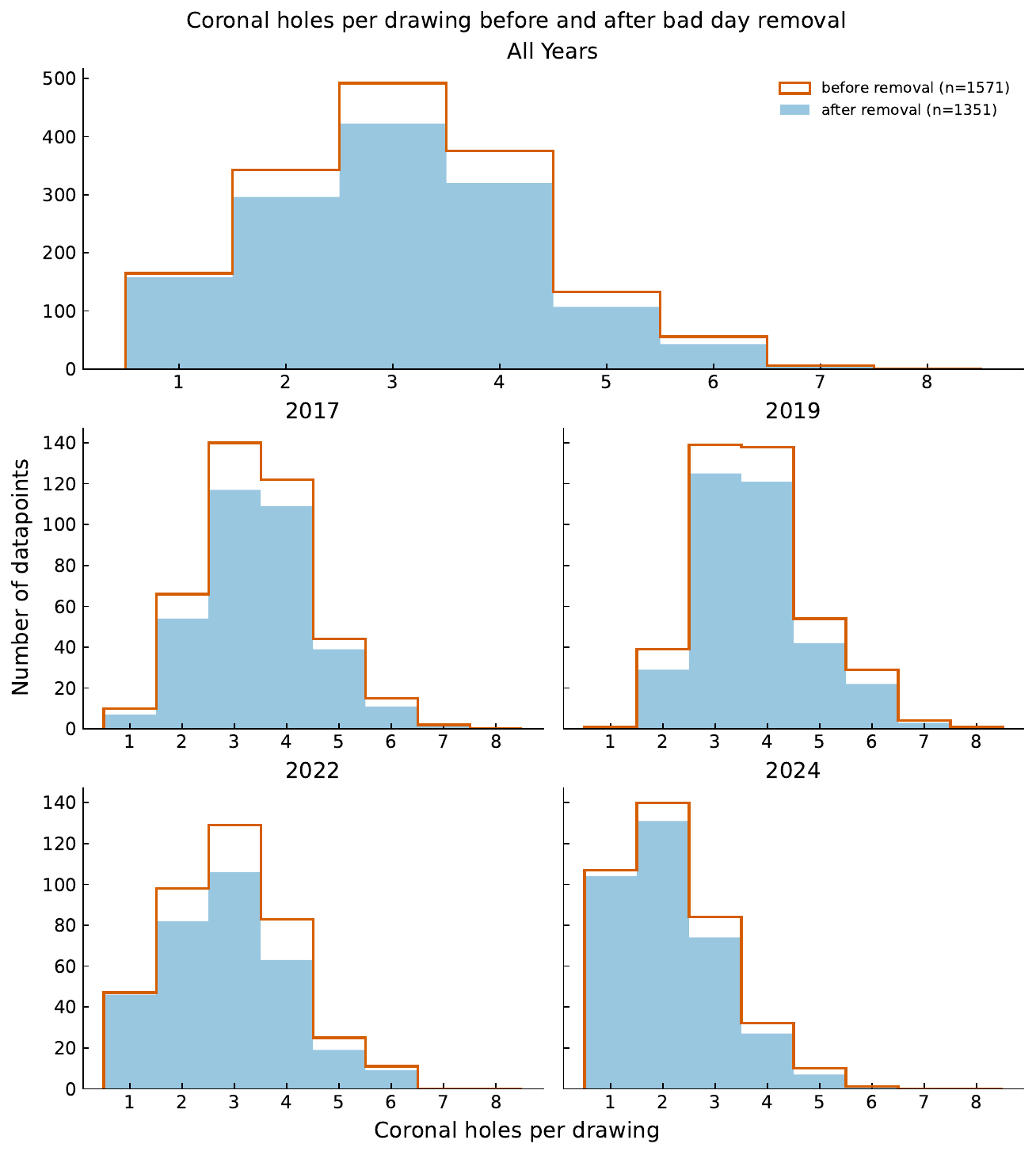}
    \caption{Distribution of CH count within the dataset before and after removal of ``bad days'' from the dataset.}
    \label{fig:CountDistributionPlot}
\end{figure}

We also report the distributions for the number of CHs each day before and after bad day removal in Figure~\ref{fig:CountDistributionPlot} and Table~\ref{tab:CountDistributionStats}. We find this also aligns with the solar cycle across the years.

\begin{table}[H]
    \caption{Coronal Hole Count Distribution Before and After Bad Day Removal}
    \label{tab:CountDistributionStats}
    \centering
    \small
    \begin{tabular*}{\linewidth}{@{\extracolsep{\fill}} l l c c c c c}
        \hline\hline
        Years & Stage & $N$ & Mean & Median & SD & Skewness \\
        \hline
        All Years & Before & 1571 & 3.105 & 3.000 & 1.266 & 0.315 \\
        All Years & After & 1351 & 3.053 & 3.000 & 1.259 & 0.282 \\
        \hline
        2017 & Before & 399 & 3.444 & 3.000 & 1.112 & 0.313 \\
        2017 & After & 339 & 3.472 & 3.000 & 1.091 & 0.299 \\
        \hline
        2019 & Before & 405 & 3.773 & 4.000 & 1.118 & 0.585 \\
        2019 & After & 343 & 3.735 & 4.000 & 1.058 & 0.562 \\
        \hline
        2022 & Before & 393 & 2.934 & 3.000 & 1.202 & 0.323 \\
        2022 & After & 325 & 2.858 & 3.000 & 1.217 & 0.367 \\
        \hline
        2024 & Before & 374 & 2.201 & 2.000 & 1.048 & 0.742 \\
        2024 & After & 344 & 2.142 & 2.000 & 1.022 & 0.784 \\
        \hline
    \end{tabular*}
    \parbox{\linewidth}{\textit{Note.} Summary statistics of the number of CHs per SWPC drawing for the full set (before) and the set retained after bad day removal (after), by year and pooled. SD is the sample standard deviation and skewness is the adjusted Fisher-Pearson coefficient.}
\end{table}


\section{Thresholding}
\label{sec:appendix_thresholding}

\par \replaced{Predictions from}{The results of the} CHRONNOS models are \replaced{pixel-wise probability images with dimensions equivalent to the input images. T}{all pixelwise values from 0-1, t}herefore, to compute binary metrics such as accuracy, IoU, \added{and} TSS, a \added{pixel value} threshold needs to be chosen. In contrast to \deleted{CHRONNOS }\citeA{jarolim_multi-channel_2021}, we choose to optimize \replaced{this threshold}{these thresholds}, which can increase performance both for the model trained on SPoCA-CH and \added{the models trained on} our CHASM-SWPC datasets\deleted{ \protect\cite{jarolim_multi-channel_2021}}. To do so, we optimized the average of the metrics \replaced{(accuracy, IoU, and TSS)}{IoU, TSS, and Accuracy}, all of which should ideally be maximized, as a simple proxy for overall performance. \added{Detection thresholds were selected via a percentile-sampled grid search of up to 1,000 candidate thresholds over each models' predicted probability distribution on the training split (January to October of each year).}

\begin{figure}[h!]
    \centering
    \includegraphics[width=\hsize]{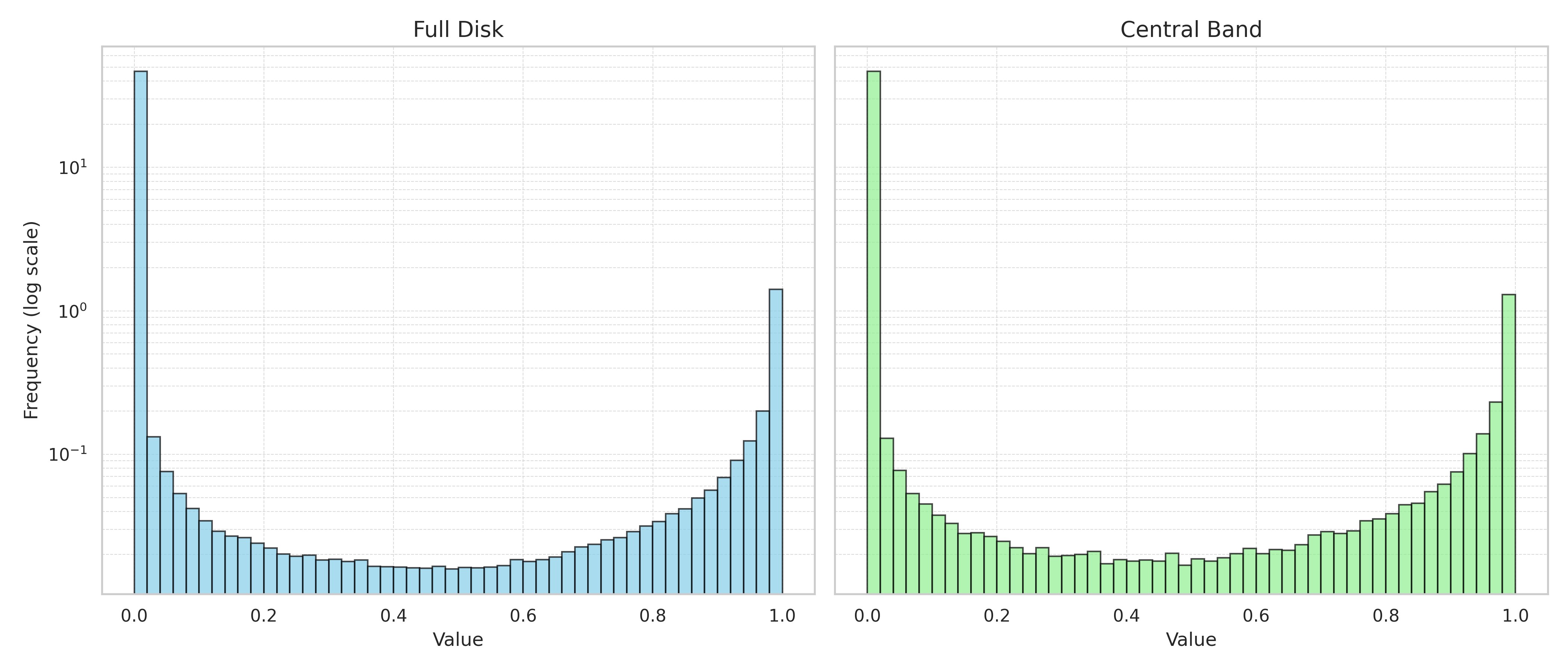}
    \caption{Histogram of values from predictions of the CHRONNOS model trained on CHASM-SWPC-1112 from 2017-01-01 through 2017-01-10 for the full solar disk and the central solar disk band, \([-400'',400'']\). 0 represents a pixel that was not classified as part of a CH, and 1 represents a pixel that was classified as part of a CH, with confidence in the prediction growing in either direction from 0.5.}
    \label{fig:pixel_distribution}
\end{figure}

\added{Based on the class imbalance between the regions with and without coronal holes, paired with the high certainty with which CHRONNOS-based models can predict them, the models tend to output a highly bimodal distribution, as shown in Figure \ref{fig:pixel_distribution}. For this reason, the optimal thresholds tend to be very high or very low as so few pixels get ``flipped'' based on a thresholding value in the middle between 0 and 1.}

\par In addition, individual metrics can be optimized, resulting in increased performance for that particular metric. Across all models, we found marginal improvements when optimizing for only one metric. Similar to the results presented in the main body (see Table \ref{tab:FullDiskMainTable} and Table \ref{tab:CentralBandMainTable}), we have prepared the results for each model when optimizing a particular threshold for each metric, see Table \ref{tab:FullDiskIndividualThresholds} and Table \ref{tab:CentralBandIndividualThresholds}.\deleted{ For more information about why the thresholds behave this way, and the motivation for thresholding, see \protect\ref{sec:appendix_result_distribution}.}

\begin{table}[h!]
    \caption{Full Disk Thresholding Results}
    \label{tab:FullDiskIndividualThresholds}
    \centering
    \small
    \begin{tabular*}{\linewidth}{@{\extracolsep{\fill}} l c c c c c c}
        \hline\hline
        Model GT & Acc. Thresh. & Acc. & TSS Thresh. & TSS & IoU Thresh. & IoU \\
        \hline
        CHASM-SWPC-111\replaced{2}{1} & 0.9654 & \textbf{0.9803} & 0.5238 & \underline{0.7793} & 0.9654 & \textbf{0.5479} \\
        CHASM-SWPC-1407 & 0.9609 & 0.9793 & 0.6773 & 0.7358 & 0.9609 & 0.5091 \\
        CHASM-SWPC-9\replaced{71}{67} & 0.9718 & \underline{0.9790} & 0.0111 & \textbf{0.8161} & 0.9443 & \underline{0.5415} \\
        SPoCA-CH & 0.7693 & 0.9759 & 0.0001 & 0.7318 & 0.3204 & 0.4875 \\
        \hline
    \end{tabular*}
    \parbox{\linewidth}{\textit{Note.} Metrics of models based on training dataset evaluated on full solar-disk of CHASM-SWPC-111\replaced{2}{1} test set, maximizing each threshold individually. \textbf{Bold} is the highest and \underline{underlined} is the second highest value in the column.}
\end{table}

\begin{table}[h!]
    \caption{Central Band Results}
    \label{tab:CentralBandIndividualThresholds}
    \centering
    \small
    \begin{tabular*}{\linewidth}{@{\extracolsep{\fill}} l c c c c c c}
        \hline\hline
        Model GT & Acc. Thresh. & Acc. & TSS Thresh. & TSS & IoU Thresh. & IoU \\
        \hline
        CHASM-SWPC-111\replaced{2}{1} & 0.9681 & \textbf{0.9665} & 0.7993 & \underline{0.7337} & 0.9540 & \textbf{0.5636} \\
        CHASM-SWPC-1407 & 0.9657 & 0.9645 & 0.7817 & 0.6917 & 0.9527 & 0.5245 \\
        CHASM-SWPC-9\replaced{71}{67} & 0.9648 & \underline{0.9647} & 0.3455 & \textbf{0.7488} & 0.9492 & \underline{0.5480} \\
        SPoCA-CH & 0.7693 & 0.8099 & 0.9589 & 0.7126 & 0.3594 & 0.5050 \\
        \hline
    \end{tabular*}
    \parbox{\linewidth}{\textit{Note.} Metrics of models based on training dataset evaluated on \replaced{central solar disk band \([-400'',400'']\)}{full solar-disk} of CHASM-SWPC-111\replaced{2}{1} test set, maximizing each threshold individually. \textbf{Bold} is the highest and \underline{underlined} is the second highest value in the column.}
\end{table}

The improvements in many of the individual metrics are marginal when compared to the metrics optimized as the average of the three metrics. With this thresholding technique, the CHASM-SWPC-111\replaced{2}{1} and the other CHASM\added{-SWPC} datasets still have significant performance benefits over SPoCA-CH. The optimal thresholds for SPoCA-CH tend to be much lower overall, and we believe this is due to the SPoCA-CH ground truth and model predictions having smaller regions than the CHASM-SWPC-111\replaced{2}{1} test set segmentations. The lower thresholds for SPoCA-CH allow more pixels to be classified as coronal holes, matching the CHASM-SWPC-111\replaced{2}{1} ground truth better.

\section{Training Parameters}
\label{sec:appendix_training_details}

Table~\ref{tab:chronnos-training-config} summarizes the training configuration for our CHRONNOS-based models. Class imbalance between CH and non-CH pixels was addressed by reweighting the negative class in the loss function by the positive:negative pixel ratio, recomputed from the training masks at each resolution stage. 

Full per-resolution batch size and epoch counts are given in Table~\ref{tab:chronnos-resolution-schedule}. 

\begin{table}[H]
    \caption{CHRONNOS Training Configuration}
    \label{tab:chronnos-training-config}
    \centering
    \small
    \begin{tabular*}{\linewidth}{@{\extracolsep{\fill}} l p{0.7\linewidth}}
        \hline\hline
        Parameter & Value \\
        \hline
        Input Channels & 8 (AIA 94, 131, 171, 193, 211, 304, 335\AA~+ HMI LoS magnetogram) \\
        Image Resolution & $8^2 \to 512^2$ px\\
        Loss Function & Per-pixel BCE, class-imbalance-reweighted \\
        Optimizer & Adam ($\beta_1{=}0$, $\beta_2{=}0.99$, weight decay $=10^{-8}$) \\
        Learning Rate & $10^{-3}$ initial; $\times 0.5$ at each new resolution stage \\
        Data Augmentation & Random horizontal/vertical flip ($p{=}0.5$ each, independent) \\
        Train / Test Split & Jan--Oct (train), Nov--Dec (test) of each year\\
        Random Seed & Not fixed; single training run per model \\
        Hardware & NVIDIA A100 GPU \\
        \hline
    \end{tabular*}
    \parbox{\linewidth}{\textit{Note.} The CHRONNOS architecture utilizes a progressive growing technique to scale the image resolution during training. The varying resolution, batch size, and epoch count for each of the 7 stages can be found in Table \ref{tab:chronnos-resolution-schedule}. No validation set was used. These parameters are unchanged from the training done by \citeA{jarolim_multi-channel_2021}.}
\end{table}

\begin{table}[H]
    \caption{Per-Resolution Training Schedule}
    \label{tab:chronnos-resolution-schedule}
    \centering
    \small
    \begin{tabular*}{\linewidth}{@{\extracolsep{\fill}} c c c}
        \hline\hline
        Resolution $\left(px\right)$ & Batch Size $\left( \frac{512}{px}\right)$ & Epochs $\left( \min\left(100,\frac{5120}{px}\right)\right)$ \\
        \hline
        $8\times8$     & 64 & 100 \\
        $16\times16$   & 32 & 100 \\
        $32\times32$   & 16 & 100 \\
        $64\times64$   & 8  & 80  \\
        $128\times128$ & 4  & 40  \\
        $256\times256$ & 2  & 20  \\
        $512\times512$ & 1  & 10  \\
        \hline
    \end{tabular*}
    \parbox{\linewidth}{\textit{Note.} Each resolution stage after the first consists of a fade-in sub-phase followed by a fixed sub-phase at that resolution. Each stage ran for the listed number of epochs. These counts are the same as those used in the work by \citeA{jarolim_multi-channel_2021}.}
\end{table}


\section{Finetuning Models}
\label{sec:appendix_finetuning}

\par We finetuned \added{CHRONNOS} models on our CHASM-SWPC datasets\added{, which were} pretrained on SPoCA-CH, to see if this would improve overall performance\deleted{, providing a starting point from the SPoCA-CH labels to train}. \replaced{Finetuning did not yield an improvement over direct training on CHASM-SWPC datasets, likely because the CHASM-SWPC datasets are comparably large when compared to the SPoCA-CH dataset.}{Likely because the CHASM-SWPC datasets are comparably large when compared to the SPoCA-CH dataset, finetuning did not yield an improvement over direct training of CHASM datasets.} We performed finetuning by taking the output of the CHRONNOS architecture, and applying a small learning rate to the final 512x512 pixel size of the progressively growing U-Net backbone of the CHRONNOS architecture \cite{jarolim_multi-channel_2021}. The results of these experiments can be found in Table \ref{tab:FullDiskFinetuning} and Table \ref{tab:CentralBandFinetuning}.

\begin{table}[h]
    \caption{Full Disk Finetuning Results}
    \label{tab:FullDiskFinetuning}
    \centering
    \small
    \begin{tabular*}{\linewidth}{@{\extracolsep{\fill}} l c c c c c}
        \hline\hline
        Model GT & FT LR & Thresh. & Acc. & TSS & IoU \\
        \hline
        CHASM-SWPC-111\replaced{2}{1} & NA & 0.5000 & 0.9740 & 0.7811 & \underline{0.5413} \\
        CHASM-SWPC-111\replaced{2}{1} & NA & 0.9328 & \textbf{0.9802} & 0.6807 &\textbf{ 0.5668} \\
        CHASM-SWPC-111\replaced{2}{1} & 1e-5 & 0.5000 & 0.9274 & \textbf{0.8200} & 0.3608 \\
        CHASM-SWPC-111\replaced{2}{1} & 1e-5 & 0.9297 & 0.9767 & 0.6942 & 0.5392 \\

        CHASM-SWPC-111\replaced{2}{1} & 1e-6 & 0.5000 & 0.9565 & 0.7322 & 0.4130 \\ 
        CHASM-SWPC-111\replaced{2}{1} & 1e-6 & 0.8860 & 0.9660 & 0.6845 & 0.4456 \\
        
        CHASM-SWPC-1407 & 1e-5 & 0.5000 & 0.9459 & \underline{0.8123} & 0.4051 \\ 
        CHASM-SWPC-1407 & 1e-5 & 0.8998 & \underline{0.9762} & 0.6922 & 0.5368 \\

        CHASM-SWPC-1407 & 1e-6 & 0.5000 & 0.9448 & 0.7759 & 0.4210 \\ 
        CHASM-SWPC-1407 & 1e-6 & 0.9886 & 0.9724 & 0.6636 & 0.4992 \\
        
        SPoCA-CH & NA & 0.5000 & 0.9759 & 0.5626 & 0.4813 \\
        \hline
    \end{tabular*}
    \parbox{\linewidth}{\textit{Note.} Metrics of models based on training dataset evaluated on full solar-disk of CHASM-111\replaced{2}{1} test set. Thresholds set at 0.5 are not optimized, other thresholds are optimized to balance average of thresholds. Finetuned learning rate (FT LR) represents the learning rate for finetuning, NA if not finetuned. \textbf{Bold} is the highest and \underline{underlined} is the second highest value in the column.}
\end{table}

\begin{table}[H]
    \caption{Central Band Finetuning Results}
    \label{tab:CentralBandFinetuning}
    \centering
    \small
    \begin{tabular*}{\linewidth}{@{\extracolsep{\fill}} l c c c c c}
        \hline\hline
        Model GT & FT LR & Thresh. & Acc. & TSS & IoU \\
        \hline
        CHASM-SWPC-111\replaced{2}{1} & NA & 0.5000 & 0.9559 & \textbf{0.7604} & \underline{0.5453}  \\
        CHASM-SWPC-111\replaced{2}{1} & NA & 0.9338 & \textbf{0.9667} & 0.6769 & \textbf{0.5707}  \\
        CHASM-SWPC-111\replaced{2}{1} & 1e-5 & 0.5000 & 0.8770 & 0.7396 & 0.3706 \\
        CHASM-SWPC-111\replaced{2}{1} & 1e-5  & 0.9412 & \underline{0.9621} & 0.6580 & 0.5404 \\
        CHASM-SWPC-111\replaced{2}{1} & 1e-6 & 0.5000 & 0.9203 & 0.7032 & 0.4131 \\
        CHASM-SWPC-111\replaced{2}{1} & 1e-6 & 0.9386 & 0.9428 & 0.6512 & 0.4538 \\
        CHASM-SWPC-1407 & 1e-5 & 0.5000 & 0.9071 & \underline{0.7511} & 0.4088 \\
        CHASM-SWPC-1407 & 1e-5 & 0.9139 & 0.9605 & 0.6550 & 0.5339 \\
        CHASM-SWPC-1407 & 1e-6 & 0.5000 & 0.8970 & 0.7180 & 0.4133 \\
        CHASM-SWPC-1407 & 1e-6 & 0.9932 & 0.9543 & 0.6290 & 0.5019 \\
        SPoCA-CH & NA & 0.5000 & 0.9592 & 0.5868 & 0.5007 \\
        SPoCA-CH & NA & 0.0112 & 0.9543 & 0.6688 & 0.5069 \\
        \hline
    \end{tabular*}
    \parbox{\linewidth}{\textit{Note.} Metrics of models based on training dataset evaluated on full solar-disk of CHASM-SWPC-111\replaced{2}{1} test set. Thresholds set at 0.5 are not optimized, other thresholds are optimized to balance average of thresholds. Finetuned learning rate (FT LR) represents the learning rate for finetuning, NA if not finetuned. \textbf{Bold} is the highest and \underline{underlined} is the second highest value in the column.}
\end{table}

%
%

\section*{Open Research Section}

The CHASM software, written in Python, is available at: \url{https://github.com/CBDevelopment/CHASM} \cite{beck2025chasm}. The code is distributed under the GNU GPLv3 license.

The CHASM-SWPC dataset used in this study, including NOAA SWPC \replaced{drawings}{synoptic maps}, Segment Anything Model (SAM) segmentations of those \replaced{drawings}{maps}, CHASM \replaced{annotations}{tool selections}, and SDO AIA and HMI imagery used in the analysis\added{ is publicly available.}\replaced{ T}{, as well as t}he trained CHASM-SWPC\deleted{-1111} model weights are \added{also} publicly available\added{.} \added{Both can be found} at Zenodo: \url{https://doi.org/10.5281/zenodo.19121464} \cite{beck2025chasm_swpc}.

\section*{Conflict of Interest disclosure}
The authors declare there are no conflicts of interest for this manuscript.

\acknowledgments
C. Beck and E. Smith contributed equally to this work. This research utilized computational resources supported by the Academic \& Research Computing group at Worcester Polytechnic Institute. We would like to thank the NOAA SWPC for their \replaced{well-}{cleanly }maintained and accessible \deleted{synoptic }drawings, and for the helpful email exchanges with members of their team. Thanks also to JSOC for their accessible supply of SDO imagery. We also appreciate the availability of the SAM from Meta.

%
%

\bibliography{chasm}

%
%
%
%
%

\end{document}